\documentclass[12pt]{iopart}
\usepackage{graphicx}
\usepackage{cite}
\usepackage{amsfonts}

\begin{document}
\topical[Non-equilibrium Phase Transitions with Long-Range Interactions]
	{Non-equilibrium Phase Transitions with Long-Range Interactions}

\author{Haye Hinrichsen}
\address{
Universit{\"a}t W{\"u}rzburg,
Fakult{\"a}t f{\"u}r Physik und Astronomie, \\
Am Hubland, 97074 W{\"u}rzburg, Germany
}
\ead{hinrichsen@physik.uni-wuerzburg.de}

\begin{abstract}
This review article gives an overview of recent progress in the field of non-equilibrium phase transitions into absorbing states with long-range interactions. It focuses on two possible types of long-range interactions. The first one is to replace nearest-neighbor couplings by unrestricted L{\'e}vy flights with a power-law distribution $P(r) \sim r^{-d-\sigma}$ controlled by an exponent~$\sigma$. Similarly, the temporal evolution can be modified by introducing waiting times $\Delta t$ between subsequent moves which are distributed algebraically as $P(\Delta t) \sim (\Delta t)^{-1-\kappa}$. It turns out that such systems with L{\'e}vy-distributed long-range interactions still exhibit a continuous phase transition with critical exponents varying continuously with $\sigma$ and/or $\kappa$ in certain ranges of the parameter space. In a field-theoretical framework such algebraically distributed long-range interactions can be accounted for by replacing the differential operators $\nabla^2$ and $\partial_t$ with fractional derivatives $\tilde{\nabla}^\sigma$ and $\tilde{\partial}_t^\kappa$. As another possibility, one may introduce algebraically decaying long-range interactions which cannot exceed the actual distance to the nearest particle. Such interactions are motivated by studies of non-equilibrium growth processes and may be interpreted as L{\'e}vy flights cut off at the actual distance to the nearest particle. In the continuum limit such truncated L{\'e}vy flights can be described to leading order by terms involving fractional powers of the density field while the differential operators remain short-ranged.
\end{abstract}

\submitto{Journal of Statistical Mechanics: Theory and Experiment}
\maketitle

\tableofcontents
\newpage
\parskip 2mm
\pagestyle{plain}


\def\rvec{{\vec{r}}}
\def\svec{{\vec{s}}}
\def\kvec{{\vec{k}}}

\section{Introduction}

In non-equilibrium statistical physics the study of phase transitions from a fluctuating active phase into absorbing configurations has been a very active field in recent years~\cite{MarroDickman99}. The primary goal of these studies is to classify continuous phase transitions by their universal properties. The most important universality class of absorbing phase transitions, which plays a paradigmatic role as the Ising model in equilibrium statistical mechanics, is Directed Percolation (DP)~\cite{Kinzel85,Hinrichsen00,Odor04,Lubeck04}. This class of phase transitions is characterized by three independent critical exponents $\beta,\nu_\perp$, and $\nu_\parallel$ which depend solely on the dimensionality of the system but not on the microscopic details of the dynamics. Directed percolation corresponds to a specific type of field theory which first appeared in the context of hadronic interactions~\cite{Moshe78,GrassbergerSundermeyer78,CardySugar80}. The structure of this field theory led Janssen and Grassberger to the so-called DP-conjecture, stating that any model that complies with a certain set of conditions should belong to the DP class~\cite{Janssen81,Grassberger82}. One of these conditions is that the interactions have to be \textit{local} in space and time.

The present review focuses on the question how the universal properties of absorbing phase transitions change if one generalized these models as to include \textit{non-local} interactions. There are several motivations to study such models. On the one hand, directed percolation is often used as a simple model for the spreading of epidemics among a population of spatially distributed individuals. In order to make the modelling more realistic, various generalizations have been introduced involving infections over long distances, incubation times, as well as other memory effects~\cite{Bailey75,Mollison77,AndersonMay91,DaleyGani99,AnderssonBritton00}. On the other hand, DP-like processes with long-range interactions seem to play a role in certain models for non-equilibrium wetting, where an interface is grown on top of an inert substrate according to dynamic rules that violate detailed balance. Finally, from a more fundamental point of view it is of theoretical interest how long-range interactions can be accounted for in a field-theoretic description of non-equilibrium phase transitions.

A possible generalization of directed percolation with a non-local transport mechanism was proposed already 40 years ago by Mollison in the context of epidemic spreading~\cite{Mollison77}. Motivated by empirical observations he suggested to consider a model where the disease is transported in random directions over long distances $r$ which are distributed according to a power law
\begin{equation}
P(r) \sim r^{-d-\sigma}\,,
\end{equation}
where $\sigma>0$ is a control exponent. In the literature such algebraically distributed long-range displacements are known as \textit{L{\'e}vy flights}~\cite{BouchaudGeorges90} and have been studied extensively in the literature, e.g. in the context of anomalous diffusion~\cite{Fogedby94a,Fogedby94b,SaichevZavslavsky97,BrockmannSokolov02}. It should be emphasized that $\sigma$ does {\it not} introduce an additional length scale, rather it changes the scaling properties of the underlying transport mechanism. It turns out that L{\'e}vy-distributed interactions in models for epidemic spreading do not destroy the transition, instead they change the critical behavior, provided that $\sigma$ is small enough~\cite{Grassberger86,MarquesFerreira94,Albano96}. More recently, Janssen {\it et al} showed that such transitions with spatial L{\'e}vy flights can be described in terms of a renormalizable field theory~\cite{JanssenEtAl99}, which allows one to compute the critical exponents to one-loop order. Moreover, this field theory predicts an additional scaling relation so that only two of the three critical exponents turn out to be independent. These results were confirmed numerically by Monte Carlo simulations in Ref.~\cite{HinrichsenHoward99}.

In the continuum limit L{\'e}vy flights can be generated by certain non-local linear operators called \textit{fractional derivatives}. Although fractional calculus is an interesting subject on its own (see e.g.~\cite{OldhamSpanier74,Jespersen99,WestEtAl03,KilbasEtAl06}) this review is not intended to discuss mathematical issues in this direction, instead it is meant as a practical guide for statistical physicists explaining how these concepts can be used in the context of absorbing phase transitions.

Likewise it is possible to introduce a similar long-range mechanism in \textit{temporal} direction. Such `temporal' L{\'e}vy flights can be understood in terms of algebraically distributed waiting times $\Delta t$ that may be interpreted in the context of epidemic spreading as incubation times between catching and passing on the infection. The probability distribution of these waiting times is given by
\begin{equation}
P(\Delta t) \sim \Delta t^{-1-\kappa}\,,
\end{equation}
where $\kappa>0$ is the temporal L{\'e}vy exponent. Unlike spatial L{\'e}vy flights, which are isotropic in space, such temporal L{\'e}vy flights are directed forward in time in order to respect causality. Recently a field-theoretic renormalization group calculation of DP with temporal L{\'e}vy was presented in Ref.~\cite{JimenezDalmaroni06}, while the mixed case of epidemic spreading by spatial L{\'e}vy flights {\it combined} with algebraically distributed incubation times was considered in Ref.~\cite{AdamekEtAl05}. As a further step towards more realistic models, Brockmann and Geisel~\cite{BrockmannGeisel03} recently studied epidemic spreading by L{\'e}vy flights on a \textit{inhomogeneous} support in the supercritical phase.

Another type of long-range interactions is motivated by models for non-equilibrium wetting, where an interface grows on top of a hard-core substrate~\cite{HinrichsenEtAl97,HinrichsenEtAl00,SantosEtAl02,HinrichsenEtAl03,GinelliEtAl05,KissingerEtAl05,RoessnerHinrichsen06}. In these models it is particularly interesting to study the effective dynamics of pinning sites at the transition from a bound to an unbound state. It turns out that in some regions of the parameter space the pinning sites can be interpreted as active sites of a directed percolation process. However, the effective rate for binding a neighboring site (corresponding to local infections in the language of DP) may depend on the actual configuration of the adjacent detached segment of the interface, giving rise to a varying rate that depends algebraically on the distance to the nearest active site. In other regions of the parameter space, where the interface fluctuates so strongly such that it may reattach far away from other pinning sites, non-local infections with power-law characteristics are observed. However, these non-local interactions differ from L{\'e}vy flights in so far as they are truncated at the actual distance of the nearest pinning site. This truncation leads to strikingly different features, as will be discussed in Sect.\ref{sec:Restricted}.

The present review is divided into three parts. The following section briefly reviews some fact about long-ranged random walks. Sect.~\ref{sec:InteractingLevy} discusses the absorbing phase transitions with unrestricted L{\'e}vy flights in space and algebraically distributed waiting times, while truncated L{\'e}vy flights are considered in Sect.~\ref{sec:Restricted}.

\section{L{\'e}vy flights, fractional derivatives, and anomalous diffusion}
\label{secAnamalous}

Before considering interacting particle systems such as directed percolation with long-range interactions, let us briefly recall some facts about long-ranged random walks by L{\'e}vy flights, also known as anomalous diffusion~\cite{MetzlerKlafter00}. We will consider two different types of long-range effects, namely, isotropic L{\'e}vy flights in space as well as algebraically distributed waiting times.

\subsection{Spatial L{\'e}vy flights}

A L{\'e}vy flight in $d$ dimensions is a random displacement $\rvec\to\rvec+\svec$ according to a probability distribution which for large distances decays algebraically as
\begin{equation}
\label{eq:SpatialLevyDistribution}
P(\svec) \sim |\svec\,|^{-d-\sigma}\,.
\end{equation}
Here the exponent $\sigma$ is a control parameter which determines the asymptotic power-law characteristics of long-range flights. Note that this parameter does not induce a length scale, instead it changes the scaling properties without breaking scale invariance. The distribution $P(\svec)$ has to be normalized by $\int {\rm d}^ds P(\svec) =1$ which implies that $\sigma>0$. Moreover, the normalization requires a lower cutoff of the power law (\ref{eq:SpatialLevyDistribution}) in the limit of small $\svec$, for example in form of a minimal flight distance.

Let us first consider a particle which moves by a series of spontaneous uncorrelated L{\'e}vy flights. The statistics of such a long-ranged random walk is described by the probability density $\rho(\rvec,t)$ to find this particle at position $\rvec$ at time~$t$, which may also be interpreted as the particle density in an ensemble of many non-interacting particles performing L{\'e}vy flights\footnote{As we will see in subsection 2.2 the interpretation of $\rho(\rvec,t)$ as a particle density is no longer valid in systems with algebraically distributed waiting times.}. Obviously, $\rho(\rvec,t)$ evolves in time according to the integro-differential equation
\begin{equation}
\frac{\partial}{\partial t} \rho(\rvec,t) \;\propto\;
\int {\rm d}^ds \, P(\svec)\, \biggl[\rho(\rvec+\svec,t)-\rho(\rvec,t)\biggr]\,.
\end{equation}
In order to write this equation in a more compact form, let us introduce a linear non-local operator $\tilde{\nabla}^{\sigma}$ which acts on a function $f(\rvec)$ by
\begin{equation}
\label{eq:SpatialLevyOperator}
\tilde{\nabla}^\sigma \,f(\rvec) = \frac{1}{\mathcal{N}_\perp(\sigma)}
\int {\rm d}^ds \, |\svec\,|^{-d-\sigma} \biggl[f(\rvec+\svec)-f(\rvec)\biggr]\,,
\end{equation}
where
\begin{equation}
\mathcal{N}_\perp(\sigma)=
-\frac{\pi^{d/2}\Gamma(-\frac{\sigma}{2})}{2^\sigma \Gamma(\frac{d+\sigma}{2})}
\end{equation}
is a $\sigma$-dependent normalization constant. Neglecting higher-order derivatives caused by the short-range cutoff, the equation of motion can be written as
\begin{equation}
\label{eq:EquationOfMotion}
\frac{\partial}{\partial t} \rho(\rvec,t) \;=\;\tilde{\nabla}^\sigma \,\rho(\rvec,t)
\end{equation}
which has the same structure as the ordinary diffusion equation apart from the fact that the Laplacian has been replaced by $\tilde{\nabla}^\sigma$.

\begin{figure}
\centering\includegraphics[width=125mm]{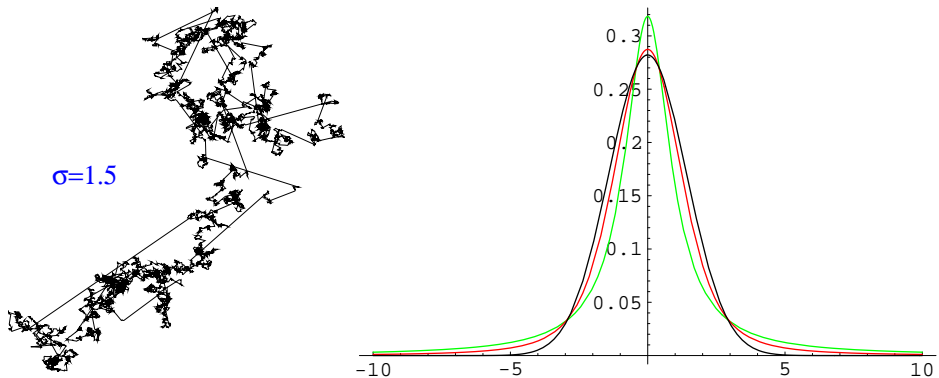}
\caption{\small
Anomalous diffusion.
Left: Trajectory of a random walk by L{\'e}vy flights for $\sigma=1.5$. Right: Gaussian distribution (black) in comparison with L{\'e}vy-stable distributions of the same width for $\sigma=1.5$~(red) and $\sigma=1$ (green).
\label{fig:distribution} }
\end{figure}

In the literature the operator $\tilde{\nabla}^\sigma$ is known as a fractional derivative because it has certain algebraic properties that generalize those of ordinary derivatives. For example, it is straight forward to show that for $0<\sigma<2$ the action of the operator $\tilde{\nabla}^\sigma$ on a plane wave amounts to bringing down a prefactor of the form
\begin{equation}
\tilde{\nabla}^\sigma \, e^{i \kvec \cdot \rvec} \;=\; -|\kvec|^\sigma \, e^{i \kvec \cdot \rvec} \,.
\end{equation}
Therefore, in momentum space the equation of motion $\frac{\partial}{\partial t} \tilde{\rho}(\kvec,t) \;=\;-|\kvec|^\sigma \,\tilde{\rho}(\kvec,t)$ implies that each mode with wave number $\kvec$ decays exponentially as $e^{-|k|^\sigma t}$ as time proceeds. Consequently, an initially localized probability distribution $\rho(\rvec,0)\propto\delta^d(\rvec)$ disperses as
\begin{equation}
\label{eq:levystable}
\rho(\rvec,t) = \frac{1}{(2\pi)^d}
\int {\rm d}^dk \,\exp\bigl(i\kvec\cdot\rvec-|\kvec|^\sigma t\bigr)\,.
\qquad (0<\sigma<2).
\end{equation}
This solution of the equation of motion~(\ref{eq:EquationOfMotion}) is called a \textit{L{\'e}vy-stable distribution}. As shown in Fig.~\ref{fig:distribution}, the tails of such a L{\'e}vy-stable distribution are more pronounced than those of an ordinary Gaussian distribution. Moreover, the width of the distribution (\ref{eq:levystable}) grows with time as $t^{1/\sigma}$, hence for $\sigma<2$ the dynamics of the random walk is superdiffusive.

The aforementioned short-range cutoff in form of a minimal flight distance generates to lowest order an additional Laplacian in the equation of motion. For $\sigma<2$ this contribution $k^2$ is less relevant than $|\kvec|^\sigma$ and can be neglected asymptotically. For $\sigma>2$, however, the $k^2$ term becomes dominant and one obtains a Gaussian distribution
\begin{equation}
\rho(\rvec,t) = \frac{1}{(4 \pi t)^ {d/2}}  \,
\exp\Bigl(-\frac{x^2}{4t}\Bigr)\,.
\end{equation}
Thus, for $\sigma>\sigma^*=2$ anomalous diffusion crosses over to ordinary diffusion.

\subsection{Algebraically distributed waiting times}

In models for ordinary diffusion particles move spontaneously to nearest-neighbor sites at a constant rate, i.e., the time intervals $\Delta t$ between subsequent moves are either constant or distributed exponentially like in Poissonian shot noise. In some situations, however, the distribution of waiting times in a random process is not exponential, instead one observes a broad distribution of waiting times. This may be described by a power-law distribution of the form
\begin{equation}
P(\Delta t) \sim \Delta t^{-1-\kappa}\,,
\end{equation}
where the exponent $\kappa$ is again a control parameter which plays a similar role as $\sigma$ in the case of spatial L{\'e}vy flights. Interpreting the waiting times $\Delta t$ as temporal displacements from event to event, they may be regarded as L{\'e}vy flights in time. However, in contrast to the spatial L{\'e}vy flights discussed above, temporal L{\'e}vy flights have to be directed forward in time ($\Delta t > 0$) in order to respect causality. Such one-sided L{\'e}vy flights differ from isotropic ones in that they are generated by a different type of fractional derivative which is non-symmetric under reflections. Its action on a Fourier mode is given by
\begin{equation}
\tilde{\partial}_t^\kappa \, e^{i\omega t} \;=\; (i\omega)^\kappa \,  e^{i\omega t} \,. \\
\end{equation}
Note that in contrast to the spatial fractional derivative $\tilde{\nabla}^\sigma$ this temporal operator brings down the nonsymmetric factor $(i\omega)^\kappa$ (rather than $-|\omega|^\kappa$) in front of the exponential, reflecting the circumstance that temporal L{\'e}vy flights are directed. For $\kappa \geq 1$ the fractional derivative defined above crosses over to its short-range counterpart $\partial_t$. The corresponding integral representation of the fractional derivative acting on a function $g(t)$ reads
\begin{equation}
\label{eq:IntegralTime}
\tilde{\partial}_t^\kappa \, g(t)  \;=\;  \frac{1}{\mathcal{N}_\parallel(\kappa)}
\int_0^{\infty} {\rm d}t' \, {t'}^{-1-\kappa} [g(t)-g(t-t')]
\qquad (0 < \kappa < 1)
\end{equation}
where
\begin{equation}
\mathcal{N}_\parallel(\kappa)=-\Gamma(-\kappa)\,
\end{equation}
is the corresponding normalization constant. In this expression the function $g(t)$ may be nonzero for all $t \in \mathbb R$, although in many applictions one uses the initial condition $g(t)=0$ for $t<0$.

The fractional derivative $\tilde{\partial}_t^\kappa$ can be used to write down an evolution equation for nearest-neighbor diffusion with waiting times
\begin{equation}
\label{eq:EquationOfMotion2}
\tilde{\partial}_t^\kappa \rho(\rvec,t) \;=\;\nabla^2 \,\rho(\rvec,t)\,.
\end{equation}
It is important to note that such a random walk with algebraically distributed waiting times is no longer Markovian. This means that the temporal evolution is no longer determined by an initial configuration at a given instance of time, instead it depends on the whole history of the process. Therefore, as a convention troughout this paper, whenever we specify an initial condition at $t=0$, we shall assume that the system was in the absorbing state for $t<0$.

For a localized initial condition at $t=0$ a formal solution of Eq.~(\ref{eq:EquationOfMotion2}) is given by
\begin{equation}
\label{eq:TemporalGreensFunction}
\rho(\rvec,t) \;=\; \frac{1}{(2\pi)^{d+1}}
\int_{-\infty}^{\infty} {\rm d} \omega \int {\rm d}^d k  \,
\frac{\exp(i \kvec \cdot \rvec - i \omega t)}{k^2-(i \omega)^\kappa}\,.
\end{equation}
Remarkably this distribution is no longer normalized as time proceeds. In fact, the integral over the whole volume
\begin{eqnarray}
N(t) &=& \int  {\rm d}^d r \rho(\rvec,t) \;=\;
\frac{1}{2\pi}
\int_{-\infty}^{\infty} {\rm d} \omega
\frac{\exp(- i \omega t)}{-(i \omega)^\kappa} \nonumber \\
&=& \frac{\Gamma(1-\kappa) \sin\pi\kappa}{\pi}\,\Theta(t)\,t^{\kappa-1}
\end{eqnarray}
decays as $N(t)\sim t^{\kappa-1}$ for $\kappa<1$. This is surprising since one expects $N(t)$ to be conserved since the particle cannot disappear. However, for processes with waiting times the quantity $\rho(\rvec,t)$ no longer describes the overall density of all random walkers, instead it just gives the density of random walkers that are about to move at the next time step $t$, excluding the waiting ones.

To explain this counterintuitive observation by an even simpler example let us consider the homogeneous equation $\tilde{\partial}_t^\kappa \,f(t)=0$. The function $f$ may be thought of as a density of non-interacting particles transported forward in time by temporal L{\'e}vy flights. Obviously, this equation is solved by $f(t)=const$, meaning that the density is conserved as time proceeds. For the corresponding Greens function, which is the solution of the inhomogeneous equation $\tilde{\partial}_t^\kappa \,f(t)=\delta(t)$, one would naively expect this conservation law to hold for $t\neq 0$. However, the solution is given by $\tilde{f}(\omega)=(2\pi)^{-1/2}(-i\omega)^{-\kappa}$, i.e.,
\begin{equation}
f(t) = \left\{
\begin{array}{ll}
const \;\; t^{\kappa-1} & \mbox{ for } t>0\\
0 &\mbox{ for } t  \leq 0
\end{array} \right.
\end{equation}
which decays in time. This decay is a consequence of the non-Markovian properties of the fractional derivative.

\subsection{Combination of spatial L{\'e}vy flights and waiting times}

\begin{figure}
\centering\includegraphics[width=65mm]{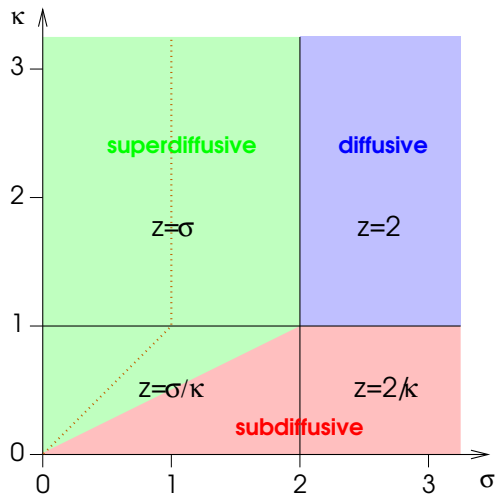}
\caption{\small
Phase diagram of anomalous diffusion. Depending on the relevancy of the L{\'e}vy operators, four different phases have to be distinguished. The background colors indicate whether the motion of the random walker is sub- or superdiffusive. The dotted line marks the boundary of ballistic motion $z=1$.
\label{fig:anomdiff} }
\end{figure}

Combining spatial L{\'e}vy flights and algebraically distributed waiting times one obtains an anomalous diffusion process with spatio-temporal long range interactions. In the continuum limit such a process is described by the partial differential equation~\cite{Fogedby94a,Fogedby94b}
\begin{equation}
 \tilde{\partial}_t^\kappa \rho(\rvec,t) \;=\; \tilde{\nabla}^\sigma \rho(\rvec,t) \,.
\end{equation}
A random walk according to this equation starting at the origin at $t=0$ is described by the formal solution
\begin{equation}
\label{eq:AnomDPGreensFunction}
\rho(\rvec,t) \;=\; \frac{1}{(2\pi)^{d+1}}
\int_{-\infty}^{\infty} {\rm d} \omega \int {\rm d}^d k  \,
\frac{\exp(i \kvec \cdot \rvec - i \omega t)}{|\kvec|^\sigma-(i \omega)^\kappa}\,.
\end{equation}
It is easy to show that this solution obeys the scaling form
\begin{equation}
\rho(\rvec,t) \;=\; t^{1-\kappa-d\kappa/\sigma}\, g\Bigl(\frac{\rvec}{t^{\kappa/\sigma}}\Bigr)
\end{equation}
with the scaling function
\begin{eqnarray}
g(\vec{z}\,) &=& g(z) = \frac{1}{(2\pi)^{d+1}}
\int_{-\infty}^{\infty} {\rm d} \omega \int {\rm d}^d k  \,
\frac{\exp(i \kvec \cdot \vec{z} - i \omega)}{|\kvec|^\sigma-(i \omega)^\kappa}\\
&=& \frac{1}{2^d \pi^{-(d+3)/2}} \,
\int_{-\infty}^{\infty} {\rm d} \omega \int_0^{\infty} {\rm d}k  \,
\int_0^{\pi} {\rm d} \theta\, \sin^{d-2}\theta \,
\frac{\exp(i k z \cos(\theta) - i \omega)}{k^\sigma-(i \omega)^\kappa}\nonumber
\,.
\end{eqnarray}
This expression is valid in the regime $\sigma<2$ and $\kappa<1$, where both fractional derivatives are relevant, while for $\sigma \geq 2$ (or $\kappa \geq 1$) the term $k^\sigma$ (or $(i \omega)^\kappa$) has to be replaced by the corresponding short-range counterpart $k^2$ (or $i\omega$). Therefore, the dynamical exponent~$z$, that relates spatial and temporal scales, is given by
\begin{equation}
z=\left\{
\begin{array}{cc}
\sigma/\kappa \quad & \mbox{ if } \sigma<2 \mbox{ and } \kappa<1 \\
\sigma\quad & \mbox{ if } \sigma<2 \mbox{ and } \kappa \geq 1 \\
2/\kappa \quad & \mbox{ if } \sigma \geq 2 \mbox{ and } \kappa<1 \\
2 \quad & \mbox{ if } \sigma \geq 2 \mbox{ and } \kappa \geq 1 \\
\end{array}\right.\,,
\end{equation}
leading to the phase diagram for anomalous diffusion shown in Fig.~\ref{fig:anomdiff}.

\subsection{Numerical simulation of L{\'e}vy flights}
\label{subsec:numerical}

In the following we describe how systems with long-range interactions can be implemented numerically. A L{\'e}vy-distributed distance vector can be created as follows. At first one generates a normalized vector equally distributed on the unit sphere. This can be done efficiently without calling time-consuming trigonometric functions by repeatedly generating a random vector in a cube until its length is smaller than~$1$ and then normalizing this vector (see code fragment in Table~\ref{tab:code}). Next one has to generate a real-valued flight distance, i.e., a floating point number $r$ distributed as
\begin{equation}
P(r)   \;=\; \left\{
\begin{array}{ll}
\sigma \, r^{-1-\sigma} & \mbox{ if } r>1 \\
0 & \mbox{ otherwise}
\end{array} \right.
\end{equation}
by setting $r:=z^{-1/\sigma}$, where $z$ is a random number drawn from a flat distribution between $0$ and~$1$. Note that $r>1$ because $1$ is the minimal flight distance. Finally the vector is multiplied by flight distance $r$. In continuum models the distance vector can be used directly while for models on a lattice it has to be rounded in such a way that the target position snaps in one of the lattice sites.

Similarly, it is possible to generate algebraically distributed waiting times by setting $\Delta t := z^{-1/\kappa}$ and, if necessary, rounding the result to a value matching the time increments of the dynamics.

\begin{table}
\begin{footnotesize}
\begin{verbatim}
                   vector GenerateLevyFlight (double sigma)
                          {
                          double x,y,z,n,r;
                          do { x=rnd(); y=rnd(); z=rnd(); } while (x*x+y*y+z*z>1);
                          n = sqrt(x*x+y*y+z*z);
                          r = pow(rnd(),-1/sigma);
                          return vector( x*r/n, y*r/n, z*r/n );
                          }
\end{verbatim}
\end{footnotesize}
\caption{\small Code fragment in C++ which generates three-dimensional isotropically distributed L{\'e}vy flights. It presupposes the existence of a suitable class for 3d vectors as well as a function {\tt rnd()} that draws a floating-point random number from a flat distribution between $0$ and~$1$. The first three lines generate a normalized and isotropically distributed vector $(x/n,y/n,z/n)$ on the unit sphere. This vector is then multiplied by the L{\'e}vy flight distance $r$.}
\label{tab:code}
\end{table}

While generating L{\'e}vy flights is simple, the real numerical challenge lies in the handling of finite size effects. Especially in those parts of the phase diagram, where the dynamics is superdiffusive, finite-size effects can be extremely dominant. There are two ways to master these difficulties:
\begin{itemize}
\item The preferred method is to eliminate finite-size effects alltogether by dynamical memory allocation. This method requires that the initial condition is localized, e.g. in form of a single particle at the origin. Instead of keeping a sparsely occupied lattice in form of a static array in the memory of the computer, the actual state of the system is stored in a dynamically generated list of particle coordinates. Suitable container classes are provided by various software packages such as the C++-Standard Template Library (STL)~\cite{StandardTemplates02}. On modern computers with 64 bit architecture, the effective lattice size of $2^{64d}$ possible coordinates can be considered as virtually infinite.
\item If a finite lattice is needed as e.g. in simulations starting with a homogeneous initial state, the proper handling of flight distances exceeding the system's boundaries is crucial. Comparing several possibilities it turned out that flights exceeding the boundaries should neither be discarded nor truncated to some maximal flight distance. Instead the cleanest scaling behavior is obtained if one applies periodic boundary conditions in their truest sense, namely, by wrapping a long-distance flight as many times around the system as needed to reach the target site. In one spatial dimension this is particularly simple since the coordinate of the target site is just $i\pm r$ modulo system size.
\end{itemize}

\subsection{Analyzing numerical simulations of systems with L{\'e}vy flights}
\label{sub:analyzing}

Monte Carlo simulations of absorbing phase transitions are usually investigated by a palette of well-established numerical methods. For example, the value of the dynamical exponent $z$ can be estimated by measuring the mean square distance from the origin $\langle r^2(t) \rangle$ which is known to grow as $t^{2/z}$. However, for systems with L{\'e}vy-type interactions this method is no longer applicable. To see this consider a single L{\'e}vy flight over distance~$r$ for $\sigma < 2$. Although the flight distance $r:=z^{-1/\sigma}$ of an \textit{individual} move is a finite number, its second moment $\langle r^2 \rangle$, i.e., the arithmetic average of the squared flight distance over an \textit{ensemble} of many independent moves, is a diverging quantity. The same applies to the mean square displacement from the origin $\langle r^2(t) \rangle$ after several subsequent moves.

As a way out, it is useful to monitor the \textit{logarithm} of the squared spreading distance and to compute the average $e^{\langle \ln(r^2(t)) \rangle}$ which amounts to study the \textit{geometric} instead of the arithmetic average. This average is finite and should grow as $t^{2/z}$ for the following reason. In systems exhibiting simple scaling (as opposed to multiscaling), the distribution of the mean square distance from the origin $r(t)$ is expected to obey the scaling form
\begin{equation}
P(\rvec,t)  \;=\; \frac{1}{t^{1/z}} \, F\bigl(\frac{\rvec}{t^{1/z}}\bigr)
\end{equation}
with some scaling function $F$. This scaling form implies that the geometric average
\begin{eqnarray}
e^{\langle \ln(r^2(t)) \rangle}  &=& \exp\left\{\int{\rm d}^d r \, P[\rvec,t] \, \ln r^2\right\} \\
&=&   \exp\left\{\int{\rm d}^d r \, F(r) \, (2 \ln r + \frac{2}{z} \ln t) \right\}
\;\propto\; \exp[\frac{2}{z} \ln t]\nonumber \\
\end{eqnarray}
grows with time as $t^{2/z}$ which allows one to determine the exponent $z$. Therefore, we can use the geometric average in the very same way as the arithmetic one is used in the case of short-range models.

\begin{figure}
\centering\includegraphics[width=155mm]{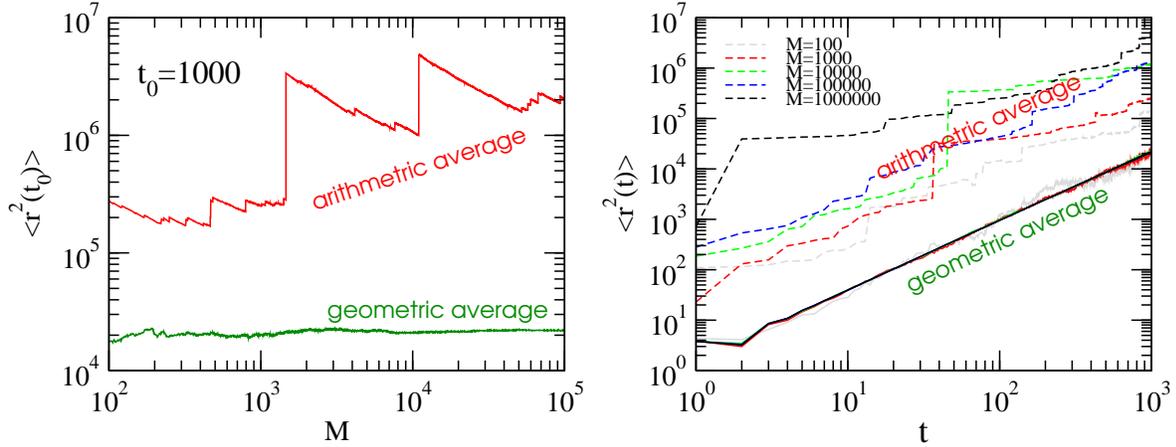}
\caption{\small
Geometric and arithmetic average of the mean square distance from the origin $r^2(t)$ of an anomalously diffusing particle in 1+1 dimensions with $\sigma=1.5$. The left panel shows the numerically computed averages after $1000$ updates depending on the ensemble size $M$. As can be seen, the geometric average converges towards a well-defined limit whereas the arithmetic average does not. The right panel shows the average for various ensemble sizes as a function of time, demonstrating that the geometric average quickly reproduces the correct power law $t^{2/\sigma}$ while the arithmetic average does not converge at all.
\label{fig:geomav} }
\end{figure}

Fig.~\ref{fig:geomav} demonstrates the strikingly different behavior of arithmetic and geometric averages in the case of anomalous diffusion for $\sigma=1.5$. As can be seen, the arithmetic average of the squared distance from the origin does not converge as the size of the statistical ensemble increases whereas the geometric average converges quickly, showing the predicted scaling behavior.

\section{Absorbing phase transitions with L{\'e}vy-distributed interactions}
\label{sec:InteractingLevy}

Having discussed anomalous diffusion of particles transported by L{\'e}vy flights in combination with algebraically distributed waiting times, we now turn to systems where particles react with each other. The simplest one to start with is the pair annihilation process $A+A\to \emptyset$ (or, equivalently, the coagulation process $A+A\to A$). Later in this section we will consider more complicated systems with phase transitions into absorbing states.

\subsection{Pair annihilation by L{\'e}vy flights}
\label{sub:pairannihilation}

In the ordinary annihilation process $A+A \rightarrow \emptyset$ with short-range interactions, the average particle density is known to decay as~\cite{Lee94}
\begin{equation}
\label{eq:annden}
\rho(t) \sim \left\{ \begin{array}{ll}
    t^{-d/2} &{\rm for \ } d < 2 \ , \\
    t^{-1} \ln t &{\rm for \ } d = d_c = 2 \ , \\
    t^{-1} &{\rm for \ } d > 2 \ .
    \end{array} \right.
\end{equation}
This leads to the question how long-interactions by L{\'e}vy flights change the dynamics, in particular the decay of the particle density and the value of the upper critical dimension. This problem was originally studied in Refs.~\cite{ZumofenKlafter94,HinrichsenHoward99}, using both simulations and approximate theoretical techniques.

\begin{figure}
\centering\includegraphics[width=85mm]{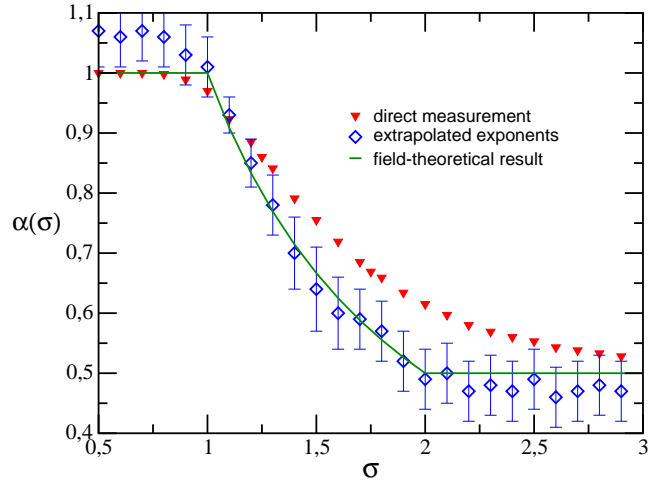}
\caption{\small
Pair annihilation $A+A \rightarrow \emptyset$ by L{\'e}vy flights. The graph shows direct estimates and extrapolations for the decay exponent $\alpha$ as a function of $\sigma$ in comparison to the exact result (solid line). Data taken from Ref.~\cite{HinrichsenHoward99}.
\label{fig:annh} }
\end{figure}

\noindent{\bf \small Numerical simulation:}\\[1mm]
%
The annihilation process $A+A \rightarrow \emptyset$ in 1+1 dimensions with L{\'e}vy-distributed interactions may be realized on a linear chain of $L$ sites with periodic boundary conditions, where each site~$i$ is either vacant ($s_i=0$) or occupied by a particle ($s_i=1$). The model evolves by random-sequential updates according to the following dynamical rules:
\begin{enumerate}
	\item Select a random site $i$. If this site is vacant ($s_i=0$) discard the update, else proceed.
	Alternatively, the site may be randomly selected from a dynamically generated list of occupied sites
	which is more efficient.
	\item Draw a random number $z \in [0,1]$ and generate a real-valued interaction distance $r := z^{-1/\sigma}$. Determine the target site $j := (i \pm \lfloor r \rfloor) \mbox{ mod } L$ in both directions with equal probability, where $\lfloor r \rfloor$ denotes truncation to the largest integer number smaller than~$r$.
	\item If the target site $j$ is empty, the particle diffuses from site $i$ to site $j$, otherwise the two particles annihilate at site $j$. Both cases can be processed by setting $s_i:=0$ and $s_j:=1-s_j$.
\end{enumerate}
Performing numerical simulations one observes that the density decays algebraically as
\begin{equation}
\label{eq:AnnhPowerLaw}
\rho(t) \sim t^{-\alpha(\sigma)}
\end{equation}
with an exponent $\alpha$ that varies continuously with the control exponent $\sigma$. Fig.~\ref{fig:annh} shows numerical estimates of $\alpha(\sigma)$ obtained in a simulation up to time $t=10^4$ on a finite lattice with $2^{16}$ sites averaged over at least $10^3$ independent runs. It turns out that the expected power law~(\ref{eq:AnnhPowerLaw}) is masked by considerable corrections to scaling, which requires to extrapolate the exponent to $t \to \infty$.\\

\noindent{\bf \small Field-theoretic analysis:}\\[1mm]
%
On the field-theoretic level pair annihilation with L{\'e}vy flights may be described by inserting the additional operator $\tilde{\nabla}^\sigma$ into the well-known field-theoretic action~\cite{Lee94}
\begin{eqnarray}
\label{AnnihilationAction}
S[\bar{\psi},\psi]
&=&
\int d^dx \, dt \,\Bigl\{
\bar{\psi} (
\partial_t - D_N \nabla^2 - D_A \tilde{\nabla}^\sigma
)\psi
\nonumber
\\ && \hspace{12mm}
+  2 \lambda \bar{\psi}\psi^2 + \lambda \bar{\psi}^2\psi^2 -
\rho_0\bar{\psi}\delta(t)
\Bigr\} \ ,
\end{eqnarray}
where $\rho_0$ is the initial homogeneous density at $t=0$. Here the field $\psi$ is {\it not} simply related to the coarse-grained density field \cite{HowardTauber97}, although the average values of both fields coincide. Note that the corresponding action for the so-called coagulation process $A+A\to A$ with L\'evy-distributed interactions would differ only in the coefficients of the reaction terms. Hence both processes are expected to have the same universal properties.

For $\sigma<2$, power counting reveals that the upper critical dimension of the model is now reduced to $d_c=\sigma<2$, hence for $d>d_c$ mean-field theory is expected to be quantitatively accurate, predicting the density decay $\rho(t)\sim t^{-1}$. Below the upper critical dimension, however, the renormalized reaction rate flows to an order $\epsilon=\sigma-d$  fixed point. Following the arguments of Ref.~\cite{Lee94}, for $d<d_c$ the only dimensionful quantity is the time $t$, which, for $\sigma<2$, scales as $[t]\sim r^{\sigma}$. Hence, for $\sigma<2$, the density decays as
\begin{equation}
\label{AnnhilationResult}
\rho(t) \sim \left\{ \begin{array}{ll}
    t^{-d/\sigma}       & {\rm for \ } d<\sigma \ , \\
    t^{-1} \ln t             & {\rm for \ } d=d_c=\sigma \ , \\
    t^{-1}              & {\rm for \ } d>\sigma \ .
    \end{array} \right.
\end{equation}
Note that for $\sigma\geq 2$ the results cross over smoothly to the standard annihilation exponents~(\ref{eq:annden}).

The field-theoretic prediction is shown in Fig.~\ref{fig:annh} as a green solid line, which is in good agreement with the extrapolated numerical results.

\subsection{Directed Percolation with L{\'e}vy flights in space}

Let us now turn to Directed Percolation (DP), which is the most important universality class of phase transitions into an absorbing state. DP is often used as a toy model for epidemic spreading: Infected individuals (active sites $A$) either infect one of their nearest neighbors ($A\to 2A$) or they recover spontaneously ($A \to \emptyset$). This competition between infection and recovery leads to a continuous phase transition out of equilibrium with universal properties.

In more realistic situations, however, short-range interactions may be inadequate to describe the transport mechanism by which the infection is spread. For example, when an infectious disease is transported by insects, their motion is typically not a random walk, rather one observes occasional flights over long distances before the next infection occurs. Similar phenomena are expected when the spreading
agent is carried by a turbulent flow. This motivated Mollison~\cite{Mollison77} to introduce a model for epidemic spreading by L\'evy flights, also called \textit{anomalous directed percolation}. It turns out that L{\'e}vy-distributed interactions do not destroy the transition, instead they may change its universal properties.\\

\noindent{\bf \small Mean field approximation and phenomenological properties:}\\[1mm]
%
Ordinary DP with short-range interactions is described by the Langevin equation~\cite{Janssen81}
\begin{equation}
\label{DPLangevin}
\tau \partial_t\rho(\rvec,t) = a \rho(\rvec,t) - b \rho^2(\rvec,t) + D \nabla^2 \rho(\rvec,t) + \xi(\rvec,t) \,,
\end{equation}
where $\nabla^2$ is the Laplacian, $\rho(\rvec,t)$ is a coarse-grained density of infected individuals, and $\tau,a,b,D$ are certain coefficients ($\tau$ is introduced for later convenience and may be set to~1). The Gaussian noise $\xi(\rvec,t)$ accounts for density fluctuations. In order to ensure the existence of an absorbing state, the noise amplitude depends on the density of active sites. According to the central limit theorem the noise is described by the correlation function
\begin{equation}
\label{Noise}
\langle \xi(\rvec,t) \xi(\rvec',t')\rangle = 2c \, \rho(\rvec,t) \, \delta^d(\rvec-\rvec')\delta(t-t')\,.
\end{equation}
Following the strategy described above, long-range infections by L{\'e}vy flights may be introduced by adding a term with a fractional derivative $\tilde{\nabla}^\sigma$ (see Eq.~(\ref{eq:SpatialLevyOperator})), leading to the modified Langevin equation
\begin{equation}
\label{eq:DPspatialLevy}
\tau \partial_t \rho =
a \rho - b \rho^2 + D \nabla^2\rho + \tilde{D}\tilde{\nabla}^{\sigma} \rho + \xi(\rvec,t)
\end{equation}
with the same noise as in Eq.~(\ref{DPLangevin}). Dimensional analysis of Eq.~(\ref{eq:DPspatialLevy}) immediately yields the mean field exponents
\begin{equation}
\label{eq:MFExponentsSpatialLevy}
\beta^{\rm MF}=1\,, \quad
\nu_\perp^{\rm MF}=\sigma^{-1}\,, \quad
\nu_\parallel^{\rm MF}=1\,, \quad
\end{equation}
and the upper critical dimension
\begin{equation}
\label{SpatialDc}
d_c = 2{\sigma}\,.
\end{equation}
As in the previous case of anomalous pair annihilation, the mean field exponents are found to vary continuously with $\sigma$ and $\kappa$. As expected, for $\sigma\to 2$ and $\kappa\to 1$ they cross over smoothly to the well-known mean field exponents of DP $\beta=\nu_\parallel=2\nu_\perp=1$ in $d \geq 4$ spatial dimensions.
%
\begin{figure}
\centering\includegraphics[width=135mm]{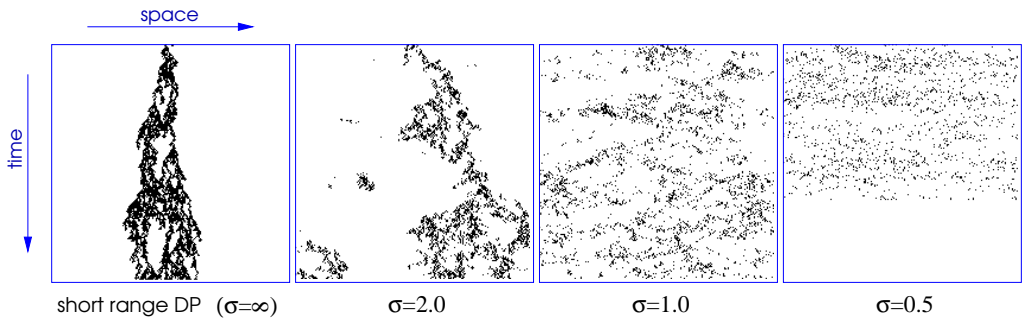}
\caption{\small
Directed percolation with spatial L{\'e}vy flights in 1+1 dimensions for various values of $\sigma$.
\label{fig:demo} }
\end{figure}
%

Spatial L{\'e}vy-distributed interactions in a DP process have two different consequences. Firstly, the clusters are no longer connected, instead the long-range interactions allow any site to become active. This is demonstrated in Fig.~\ref{fig:demo}, where the spatio-temporal evolution of a critical 1+1-dimensional process is shown for various values of $\sigma$. Secondly, long-range interactions tend to enhance diffusive mixing, leading to a more homogeneous impression of the particle density as $\sigma$ decreases. Accordingly the upper critical dimension decreases since long-range interactions tend to diminish the influence of fluctuations.

In the rightmost panel of Fig.~\ref{fig:demo} the process suddenly enters the absorbing state. Such finite-size effects are typical for phase transitions into absorbing states with long-range interactions. Within mean field theory, finite-size effects are expected to set at a typical time that scales as $L^\sigma$, where $L$ is the lateral system size. Thus, especially for small values of $\sigma$ finite-size effects can be extremely dominant.\\

\noindent{\bf \small Field-theoretic renormalization group calculation:}\\[1mm]
%
Below the upper critical dimension $d<d_c$ directed percolation with L{\'e}vy flights is expected to be dominated by fluctuation effects. Here the critical exponents are no longer given by the mean field values~(\ref{eq:MFExponentsSpatialLevy}), instead they have to be computed by means of a field-theoretic renormalization group calculation.

\begin{figure}
\centering\includegraphics[width=105mm]{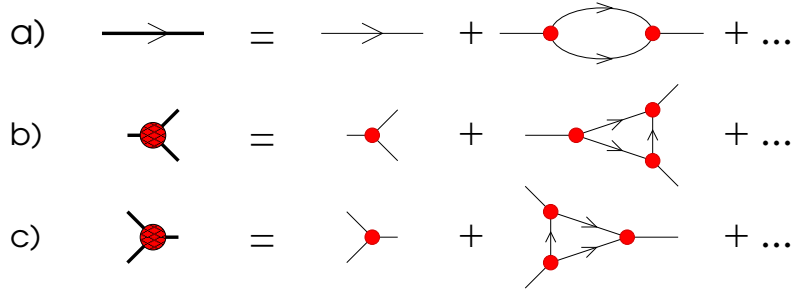}
\caption{\small
One-loop Feynman diagrams of a) the propagator and b)-c) the cubic vertices of directed percolation. The thin lines represent the free propagator $G_0(\kvec,\omega)$ directed in time according to the arrow. The small circles denote the symmetrized bare cubic vertices $\pm g$. L{\'e}vy-distributed interactions modify the free propagator but they do not change the structure of the loop expansion.
\label{fig:feynman} }
\end{figure}

Starting point of a field-theoretic renormalization group calculation is an effective action~$S$ which is defined as the partition sum over all realizations of the density field $\psi(\rvec,t):=\rho(\rvec,t)$ and the noise field $\xi(\rvec,t)$, constrained to solutions of the Langevin equation~(\ref{eq:DPspatialLevy}). Integrating out the noise by introducing a Martin-Siggia-Rosen response field $\bar{\psi}(\rvec,t)$ and symmetrizing the coefficients of the cubic terms by rescaling the fields $\psi$ and $\bar{\psi}$ one arrives at the field-theoretic action $S[\psi,\bar{\psi}]=S_0[\psi,\bar{\psi}]+S_{\rm int}[\psi,\bar{\psi}]$ consisting of a free contribution
\begin{equation}
S_0[\psi,\bar{\psi}]\,=\,\int\,{\rm d}^dx\,{\rm d}t \,\, \bar{\psi}
(\tau \partial_t 
-a-D\nabla^2 -\tilde{D} \tilde{\nabla}^{\sigma})
\psi
\end{equation}
and an interaction part
\begin{equation}
S_{\rm int}[\psi,\bar{\psi}]\,=\,g\,\int\,{\rm d}^dx\,{\rm d}t \,\,
(\bar{\psi}\psi^2-\bar{\psi}^2\psi)
\end{equation}
with the coupling constant $g=\sqrt{bc}$. Note that the additional fractional derivative $\tilde{\nabla}^{\sigma}$ appears only in the \textit{free} part of the action. Therefore, the structure of the Feynman diagrams in a loop expansion is the same as in ordinary DP (see Fig.~\ref{fig:feynman}), the only difference being that the free propagator in momentum space $G_0^{\rm DP}(\kvec,\omega) =(D k^2-a-i\tau\omega)^{-1}$ is replaced by the generalized counterpart
\begin{equation}
G_0(\kvec,\omega) = \Bigl[ D k^2 + \tilde{D}k^\sigma -a-i\tau \omega
\Bigr]^{-1} \,.
\end{equation}
%

\noindent
Apart from this modification the loop integrals have exactly the same structure and the same multiplicities as in DP. In particular, the so-called rapidity reversal symmetry
\begin{equation}
\psi(\kvec,\omega)\to-\bar{\psi}(\kvec,-\omega)\,\qquad
\bar{\psi}(\kvec,\omega)\to-\psi(\kvec,-\omega)
\end{equation}
is still valid. As a consequence the fields $\psi$ and $\bar{\psi}$ scale identically and the cubic vertices shown in Fig.~\ref{fig:feynman}b-\ref{fig:feynman}c renormalize exactly in the same way.

As first shown by Janssen \etal~\cite{JanssenEtAl99}, fractional derivatives used in interacting field theories have unexpected properties. Firstly, the loop corrections do not renormalize the L{\'e}vy operator $\tilde{\nabla}^\sigma$, instead they generate short-range contributions, the most relevant being proportional to $k^2$. This is the reason why the Laplacian has to be retained in the field-theoretic action for $d\leq d_c$. In fact, even if this term was not included, it would be generated under renormalization group (RG) transformations. As shown in the Appendix, the circumstance that the fractional derivative is not renormalized induces an additional scaling relation
\begin{equation}
\label{eq:SpatialScalingRelation}
\nu_\parallel-\nu_\perp(\sigma-d)-2\beta = 0\,
\end{equation}
among the three standard exponents $\beta$, $\nu_\perp$, and $\nu_\parallel$. The second observation by Janssen \etal~\cite{JanssenEtAl99} is even more important. As in the previous examples of anomalous diffusion and annihilation, one would expect L{\'e}vy-DP to cross over to ordinary DP at $\sigma=2$ since this is the threshold
from where on $\tilde{\nabla}^\sigma$ behaves in the same way as a Laplacian $\nabla^2$. However, Janssen \etal showed that this naive expectation is no longer correct in an interacting field theory:
\begin{quote}
\small
\textit{``Because the anomalous susceptibility exponent, the analog of the Fisher exponent in critical equilibrium phenomena, is negative here, the continuous crossover between long-range and short-range behavior arises at a L{\'e}vy exponent} [$\sigma$] \textit{greater than 2.''}
\end{quote}
This means that in an interacting field theory the L{\'e}vy operator $\tilde{\nabla}^{\sigma}$ does \textit{not} cross over to $\nabla^2$ at $\sigma=2$, rather its long-range nature prevails even beyond this threshold up to some $\sigma^*>2$. Assuming a continuous crossover of the critical exponents the threshold $\sigma^*$ can be computed by plugging the numerically known critical exponents of short-range DP into the scaling relation~(\ref{eq:SpatialScalingRelation}) and to solve it for $\sigma$, giving
\begin{equation}
\sigma^* = d+z-2\beta/\nu_\perp\,
\end{equation}
with $\sigma^*=2.0766(2)$ in one, $\sigma^*\simeq 2.2$ in two, and $\sigma^*=2+\tilde\epsilon/12$ in $d=4-\tilde\epsilon$ spatial dimensions. \\

\noindent{\bf \small Numerical results:}\\[1mm]
%
\begin{figure}
\centering\includegraphics[width=105mm]{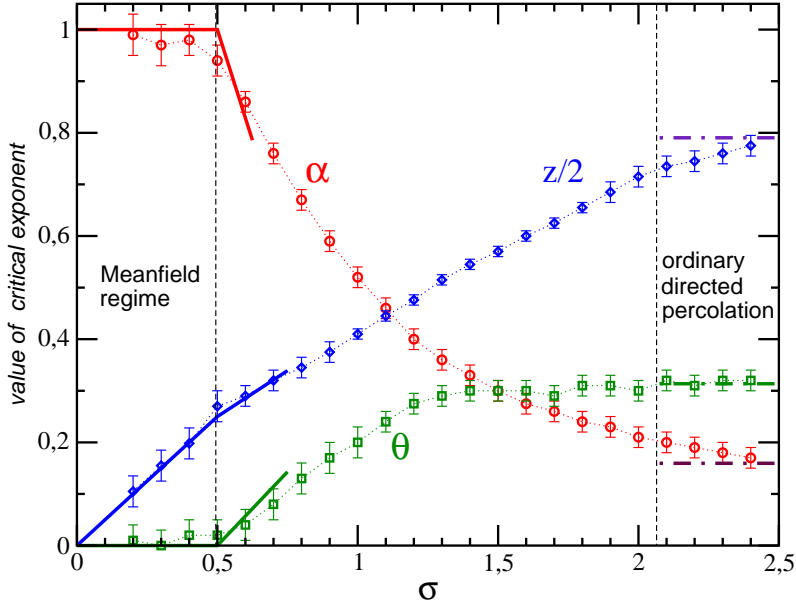}
\caption{\small
Numerically determined critical exponents of directed percolation with spatial L{\'e}vy flights. The solid lines indicate the mean field predictions and the field-theoretic results to one-loop order (data taken from~\cite{HinrichsenHoward99}).}
\label{fig:spatial}
\end{figure}
%
As shown in~\cite{JanssenEtAl99} (see also Appendix) a field-theoretical renormalization group calculation to one-loop order for DP with L{\'e}vy flights yields the critical exponents
\begin{equation}
\label{eq:SpatialCriticalExponents}
\beta = 1-\frac{2\epsilon}{7\sigma} + O\left( \epsilon^{2}\right)
,\quad
\nu_{\perp} = \frac{1}{\sigma} + \frac{2\epsilon}{7\sigma^2}
 + O\left( \epsilon^{2}\right)
,\quad
\nu_{\parallel} = 1 + \frac{\epsilon}{7\sigma} + O\left( \epsilon^{2}\right).
\end{equation}
In order to verify these predictions numerically, it is advisable to perform epidemic simulations starting with a single seed~\cite{GrassbergerTorre79} using a dynamically generated list of active sites. As described in Sect.~\ref{subsec:numerical} this simulation technique virtually eliminates finite size effects, which would be very strong especially when $\sigma$ is small. As usual in this type of simulations, one measures the survival probability $P_s(t)\sim t^{-\delta}$, the average number of particles $N(t)\sim t^\theta$, and the mean square spreading from the origin $R^2(t)\sim t^{2/z}$. Since the latter diverges in the presence of L{\'e}vy flights, the arithmetic average has to be replaced by a geometric average, as explained in Sect.~\ref{sub:analyzing}. Having determined the critical point for each value of $\sigma$ separately, the power law decay of these quantities allows one to determine the exponents $\alpha=\delta=\beta/\nu_\parallel$, $z=\nu_\parallel/\nu_\perp$, and $\theta=\frac{d}{z} - 2\beta/\nu_\parallel$.

In Fig.~\ref{fig:spatial} the analytical predictions are compared with the numerical estimates of the critical exponents in 1+1 dimensions. For $\sigma<1/2$ very long L{\'e}vy flights are so dominant that they wash out any fluctuations, hence the system is effectively described by mean field exponents. In an intermediate regime $1/2 < \sigma < \sigma^*$ the exponents vary continuously. Here the field-theoretic one-loop results correspond to the tangents (solid lines) at the left boundary, being in qualitative agreement with the numerical findings. Finally,
for $\sigma > \sigma^*$ the numerical results confirm the crossover to ordinary DP, although the results are not precise enough to verify the predicted value of $\sigma^*$. This example shows that the strength of field theory is not in a quantitative prediction of the exponents, rather the field-theoretic approach provides a conceptual understanding of the underlying process, allows one to derive exact scaling relations, and explains the origin of universality.

\subsection{Directed Percolation with waiting times}

As in the case of pure diffusion, it is straight-forward to generalize ordinary DP as to include algebraically distributed waiting times. The corresponding Langevin equation has now a fractional derivative in time:
\begin{equation}
\tilde{\partial}^\kappa_t \rho + \partial_t \rho =
a \rho - b \rho^2 + D \nabla^2\rho + \xi(\rvec,t)\,.
\end{equation}
As before, we keep the operator $\partial_t \rho$ which reflects the short-range cutoff of the distribution. In analogy to spatial L{\'e}vy flights, it is also natural to expect that such a term will be generated under renormalization group transformations. Ignoring fluctuation effects, dimensional analysis yields the mean field exponents
\begin{equation}
\label{MFExponentsTemporal}
\beta^{\rm MF}=1\,, \quad
\nu_\perp^{\rm MF}=1/2\,, \quad
\nu_\parallel^{\rm MF}=\kappa^{-1}\,, \quad
\end{equation}
and the upper critical dimension $d_c = 6-2/\kappa$. Note that it is now the temporal exponent $\nu_\parallel$ that varies continuously with $\kappa$. Like spatial L{\'e}vy flights, algebraically distributed waiting times decrease the upper critical dimension, meaning that they tend to wash out fluctuation effects.

Using similar arguments as in the case of spatial L{\'e}vy-distributed interactions one can show that the fractional operator $\tilde{\partial}_t^\kappa$ does not renormalize itself, instead it renders corrections of the form $i\omega$ and hence renormalizes the short-range operator $\partial_t$. This leads to the scaling relation analogous to Eq.~(\ref{eq:SpatialScalingRelation}):
\begin{equation}
\label{eq:TemporalScalingRelation}
\nu_\parallel(1-\kappa)+d\nu_\perp-2\beta = 0\,.
\end{equation}
Recently Jim\'enez-Dalmaroni~\cite{JimenezDalmaroni06} presented a one-loop renormalzation group calculation, reporting that the critical exponents in $d=d_c-\epsilon$ dimensions are given by
\begin{eqnarray}
\beta &=& 1+\biggl(\frac{8(2 \kappa -1)}{\tan(\pi\kappa/2)}-[F_1(\kappa)+F_2(\kappa)]
\frac{3\kappa-1}{2} - \frac{\kappa F_3(\kappa)}{4} \biggr)
\,\frac{\epsilon}{\kappa F_3(\kappa)}+  O(\epsilon^2) \nonumber\\
\nu_\perp &=& \frac{1}{2}+\bigg(\nonumber \frac{8(2\kappa-1)}{\kappa\tan(\frac{\pi}{2}\kappa)}-F_1(\kappa)-F_2(\kappa)\biggr)
\frac{\epsilon}{2F_3(\kappa)}+O(\epsilon^2).\\
\nu_\parallel&=&\frac{1}{\kappa}+\frac{8 (2\kappa-1)}{\kappa^2 F_3(\kappa)\tan(\frac{\pi}{2}\kappa)}\epsilon+O(\epsilon^2).
\end{eqnarray}
where
\begin{eqnarray}
F_{1}(\kappa)&=&2 \biggl(2-\frac{1}{\kappa}\biggr)\frac{\sin(\pi\kappa)}{\sin^2(\frac{\pi}{2}\kappa)}\nonumber\\
F_{2}(\kappa)&=&\frac{1}{\kappa\sin^{3}(\frac{\pi}{2}\kappa)}\biggl[\cos\biggl(\frac{\pi}{2}\kappa\biggr)+\frac{4\kappa^2-3\kappa+1}{3\kappa-1}\cos\biggl(\frac{3\pi}{2}\kappa\biggr)\biggr]\\
F_3(\kappa)&=&(32-{d_c}) F_1(\kappa)+d_{c} F_{2}(\kappa).\nonumber
\end{eqnarray}
%

\subsection{Directed Percolation with spatio-temporal long-range interactions}

Finally it is possible to consider the mixed case of epidemic spreading by spatial L{\'e}vy flights {\it combined} with algebraically distributed waiting times. Power counting in the corresponding Langevin equation
\begin{equation}
\label{eq:FullLangevin}
\tilde{\tau} \tilde{\partial}_t^\kappa \rho =
a \rho - b \rho^2 + \tilde{D}\tilde{\nabla}^{\sigma} \rho + \xi(\rvec,t)
\end{equation}
with the same noise as in Eq.~(\ref{Noise}) yields the mean field exponents
\begin{equation}
\label{MFExponentsSpatioTemporal}
\beta^{\rm MF}=1\,, \quad
\nu_\perp^{\rm MF}=\sigma^{-1}\,, \quad
\nu_\parallel^{\rm MF}=\kappa^{-1}\, \quad
\end{equation}
and the upper critical dimension
\begin{equation}
\label{dcmixed}
d_c = 3\sigma - \frac{\sigma}{\kappa}\,.
\end{equation}
In particular, the dynamical exponent locks onto the ratio $z=\sigma/\kappa$.

\begin{figure}
\centering\includegraphics[width=125mm]{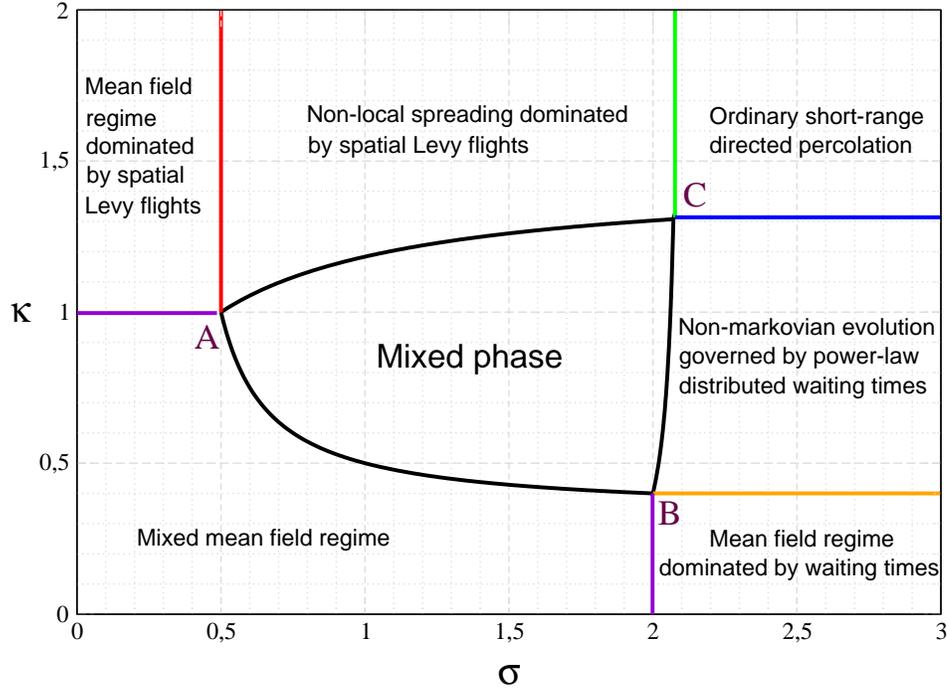}
\caption{\small
Phase diagram for epidemic spreading with spatio-temporal long-range interactions in one dimension. The seven phases and their boundaries are explained in the text.
\label{fig:phasediag} }
\end{figure}

A field-theoretic approach to the mixed case can be found in~\cite{AdamekEtAl05}. It turns out that in this case both scaling relations~(\ref{eq:SpatialScalingRelation}) and~(\ref{eq:TemporalScalingRelation}) are simultaneously valid, hence only one of the three exponents is independent. To one-loop order one finds the results
\begin{equation}
\label{GeneralOneLoopResults}
\beta = 1 - \frac{\epsilon}{6}+{\rm O}(\epsilon^2),\quad
\nu_\parallel = 1 + \frac{\epsilon}{12}+{\rm O}(\epsilon^2) ,\quad
\nu_\perp = \frac12+\frac{\epsilon}{16}+{\rm O}(\epsilon^2) .
\end{equation}
The phase diagram of a 1+1-dimensional DP process with spatio-temporal long-range interactions controlled by $\sigma$ and $\kappa$ is sketched in Fig.~\ref{fig:phasediag}. It comprises seven different phases:
\begin{enumerate}
	\item a mixed mean field phase for small $\sigma$ and $\kappa$ which is governed both by L{\'e}vy flights \textit{and} waiting times. In this phase fluctuation effects are irrelevant and the exponents are given by Eq.~(\ref{MFExponentsSpatioTemporal}).

	\item a mean field phase for $\kappa>1$ governed solely by spatial L{\'e}vy flights. In this regime waiting times described by the operator $\tilde{\partial}_t^\kappa$ are irrelevant compared to $\partial_t$ and the exponents are given by Eq.~(\ref{eq:MFExponentsSpatialLevy}).

	\item a mean field phase for $\sigma>2$ governed solely by waiting times. Here spatial L{\'e}vy flights associated with the operator $\tilde{\nabla}^{\sigma}$ become irrelevant and the exponents are given by Eq.~(\ref{MFExponentsTemporal}).

	\item a non-trivial fluctuating mixed phase in the center, where spatial L{\'e}vy flights and incubation times are both relevant. Here both scaling relations~(\ref{eq:SpatialScalingRelation}) and~(\ref{eq:TemporalScalingRelation}) are simultaneously valid, the upper critical dimension is $d_c = 3\sigma - \frac{\sigma}{\kappa}$,  and the exponents are given to one-loop order by Eq.~(\ref{GeneralOneLoopResults}).

	\item a non-trivial fluctuating phase dominated exclusively by spatial L{\'e}vy flights, where incubation times are still irrelevant so that the results of Refs.~\cite{JanssenEtAl99,HinrichsenHoward99} apply, see Eqs.~(\ref{SpatialDc}), ~(\ref{eq:SpatialScalingRelation}) and ~(\ref{eq:SpatialCriticalExponents}).

	\item a non-trivial fluctuating phase dominated by incubation times, where spatial  L{\'e}vy flights are irrelevant so that the results of Ref.~\cite{JimenezDalmaroni06} apply, see above.

	\item a short-range regime where the process behaves effectively in the same way as an ordinary DP process. In 1+1 dimensions the corresponding exponents are given by
	$\beta^{\rm DP} \simeq 0.2765$, $\nu_\perp^{\rm DP}\simeq 1.097$, and $\nu_\parallel^{\rm DP}\simeq 1.734$.
\end{enumerate}

\subsection{Synchronization of extended chaotic systems with L{\'e}vy-type interactions}

Long-range interactions by L{\'e}vy flights can also be studied in various other contexts. For example, in a recent paper Tessone \etal~\cite{TessoneEtAl06} studied a coupled map lattice defined by the update rule
\begin{eqnarray}
\label{eq:CML}
x_i(t+1) &=& (1-\gamma)\,F(\tilde{x}_i(t))+\gamma F(\tilde{y}_i(t))   \\
y_i(t+1) &=& (1-\gamma)\,F(\tilde{y}_i(t))+\gamma F(\tilde{x}_i(t)) \,.  \nonumber
\end{eqnarray}
Here $t$ and $i$ are discrete temporal and spatial indices while $x_i(t),y_i(t)$ are the state variables of two linear chains of coupled maps with periodic boundary conditions. The map $F$ is chaotic; a typical choice would be the Bernoulli map $F(x)=(2x) \,{\rm mod }\, 1$. In addition, the two chains are coupled crosswise controlled by the parameter $\gamma$. Depending on $\gamma$ the two chains either synchronize or they evolve into independent chaotic states. The transition from independent dynamics to synchronization is characterized by the Hamming distance and can be interpreted as a transition from a fluctuating phase into an absorbing state.

In the update rules the functions $\tilde{x}_i(t)$ and $\tilde{y}_i(t)$ are linear combinations of the state variables and are used to encode different coupling schemes. For example, ordinary short-range interactions would correspond to $\tilde{z}_i(t)=(1-\epsilon)z_i+\epsilon(z_{i-1}+z_{i+1})$. Obviously the generalized coupling scheme for non-local L{\'e}vy-distributed interactions would be given by
\begin{equation}
\label{eq:fullycoupled}
\tilde{z}_i(t) \;=\; (1-\epsilon)z_i+\frac{\epsilon}{\eta(\sigma)}\,
\sum_{j>0} \frac{z_{i-j}+z_{i+j}}{j^{1+\sigma}}\,,
\end{equation}
where $j$ is the interaction distance with a suitable cutoff at the system size and $\eta(\sigma)$ is the corresponding normalization. However, using the language of DP, this update algorithm first selects a non-infected site and looks from where an infection could come, thus summing over all sites of the chain. Consequently, each of these updates requires $\mathcal{O}(L^2)$ operations which strongly limits the numerically manageable system sizes. Contrarily, the update rules discussed in the previous section determine the target site to where the infection goes which takes only $\mathcal{O}(L)$ operations.

To circumvent this problem Tessone \etal present a remarkable approach which could be pathbreaking for many similar problems. The idea is to replace the fully-coupled interaction scheme of Eq.~(\ref{eq:fullycoupled}) by a \textit{diluted} coupling scheme which has nethertheless the same scaling properties. More specifically, Tessone \etal restrict the couplings to a sparse set of allowed interaction ranges
\begin{equation}
j_m(q) = q^m-1\,,
\end{equation}
where $q$ is a parameter with values typically chosen as $q=2$,$4$, and $8$. The index $m$ enumerates the allowed interaction distances and runs from $1$ to $M=\log_q(L/2)$. The coupling scheme is then given by
\begin{equation}
\tilde{z}_i  \;=\; (1-\epsilon)z_i +
\frac{\epsilon}{\eta(\sigma)}\,\sum_{m=1}^M \frac{z_{i-j_m(q)}+z_{i+j_m(q)}}{[j_m(q)]^\sigma}\,,
\end{equation}
which differs from the fully-coupled one in Eq.~(\ref{eq:fullycoupled}) in so far as the interaction distance in the denominator is raised to the power $\sigma$ instead of $1+\sigma$. As shown Tessone et al. this modification ensures that the diluted coupling scheme has effectively the same statistical weight as in the fully-coupled case. Because of the dilution the computational effort per site reduces to $\mathcal{O}(L \log_q L)$ operations per update which allows one to simulate much larger systems.

Tessone \etal studied the scaling behavior of the synchronization error $w_i(t)=|x_i(t)-y_i(t)|$ in simulations of a coupled map lattice with up to $4 \times 10^6$ maps (sites). Comparing their numerical results with those of Ref.~\cite{HinrichsenHoward99} they find that the system belongs to the same universality class as anomalous directed percolation with spatial L{\'e}vy-distributed interactions.

\subsection{Parity-conserving processes with spatial L{\'e}vy flights}

Although so far most papers focused on generalizations of directed percolation, the concept of long-range interactions can be applied to any system with a phase transition into absorbing states. Another important universality class of such phase transitions is the so-called parity-conserving (PC) class~\cite{GrassbergerEtAl84,TakayasuTretyakov92,CardyTauber96,Hinrichsen97} in 1+1 dimensions, sometimes also referred to as the voter universality class~\cite{DornicEtAl01}. This process is described by the reaction-diffusion scheme
\begin{equation}
2A \to 3A \,, \qquad 2A \to \emptyset
\end{equation}
and exhibits a non-equilibrium phase transition from a fluctuating phase with a stationary density of particles to an absorbing phase dominated by pair annihilation with an algebraic decay $\sim t^{-1/2}$. The upper critical dimension of this process, above which mean-field behavior prevails, is $d_c=2$, while in an intermediate regime $d_c'<d<d_c$ with $d_c'=4/3$ the dynamics is governed by the annihilation fixed point~\cite{Lee94} with a zero branching rate at the critical point, see Eq.~(\ref{eq:annden}). Only below $d_c'$ the critical branching rate becomes nonzero and the transition is characterized by non-trivial properties. Unfortunately, this case cannot be described by a perturbative renormalization group calculation\cite{CardyTauber96}.

\begin{figure}
\centering\includegraphics[width=65mm]{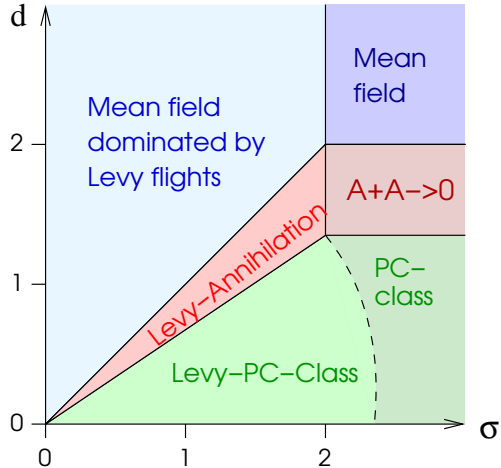}
\caption{\small
General phase structure of parity-conserving branching annihilating random walks with L{\'e}vy-distributed long range interactions.
\label{fig:levybarw} }
\end{figure}

It is now straight-forward to replace nearest-neighbor couplings by non-local interactions according to a symmetric L{\'e}vy distribution controlled by $\sigma$. In a field-theoretical and numerical study in 1+1 dimensions Vernon and Howard~\cite{VernonHoward01} observed the following phase structure (see Fig.~\ref{fig:levybarw}):
\begin{enumerate}
\item[(a)] For $\sigma>2.5$ the process is effectively a short-ranged and the transition belongs to the usual PC class.
\item[(b)] In a range $1.5<\sigma<2.5$ one observes a nontrivial phase with continuously varying exponents.
\item[(c)] For $\sigma<1.5$ the critical branching rate is zero and one finds the same behavior as for pair annihilation with L{\'e}vy flights as described in Sect.~\ref{sub:pairannihilation}, where the density decay according to Eq.~(\ref{AnnhilationResult}).
\item[(d)] For $\sigma<1$ the process displays mean field behavior with the density decaying away as $\rho\sim 1/t$ at zero branching rate.
\end{enumerate}

The observed phenomenology in 1+1 dimensions suggests the general phase structure shown in Fig.~\ref{fig:levybarw}. Vernon and Howard show that the phase boundary, where the branching rate becomes zero, is given by $\sigma_c'(d) = 3d/2$, while the precise form of the phase boundary between short- and long-ranged branching annihilating random walks (the dashed line in Fig.~\ref{fig:levybarw}) is not yet known. However, taking field-theoretic action given in Ref.~\cite{VernonHoward01} and assuming that the fractional derivative is not renormalized, it is near at hand to conjecture the scaling form
\begin{equation}
\sigma = d + z (1-\alpha-\delta)
\end{equation}
with $\delta=0$. This relation should hold in the phases (b) and (c), where the exponents vary continuously with $\sigma$, and gives the correct phase boundary of the annihilation-dominated regime. If this relation were correct, it would give the phase boundary of phase (b) (the dashed line) by plugging in the numerically known exponents of the PC class and solving the equation for $\sigma$. In one dimension this would give the threshold $\sigma_c \approx 2.50$, which is in qualitative agreement with the numerical simulations of Ref.~\cite{VernonHoward01}.

\section{Restricted long-range interactions}
\label{sec:Restricted}

Recently phase transitions into absorbing states with long-range interactions emerged in a different context, namely, in wetting processes out of equilibrium~\cite{HinrichsenEtAl97,HinrichsenEtAl00,SantosEtAl02,HinrichsenEtAl03,GinelliEtAl05,KissingerEtAl05,RoessnerHinrichsen06}. Here one studies an interface growing on a $d$-dimensional hard-core substrate. The interface evolves in time by deposition and evaporation processes which generally violate detailed balance. Depending on their rates the interface is either pinned to the substrate or it detaches, leading to a so-called wetting transition.

\begin{figure}
\centering\includegraphics[width=125mm]{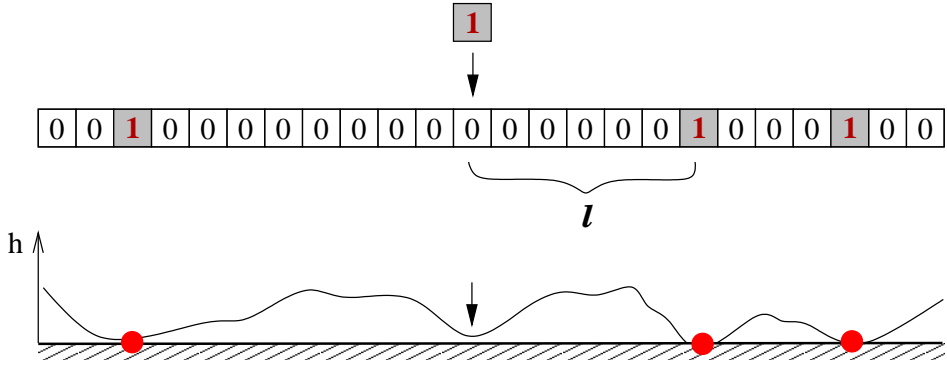}
\caption{\small
Non-equilibrium wetting process close to the unbinding transition. The contact points, where the interface touches the substrate (red points), may be interpreted as active sites of a directed percolation process. Figure taken from~\cite{GinelliEtAl05}.
\label{fig:reattach}}
\end{figure}

A schematic typical configuration of such an interface close to the wetting transition is shown in Fig.~\ref{fig:reattach}. It consists of detached domains separated by pinned segments. The dynamics of the interface is such that each detached domain can shrink or expand from its edges. New domains may be created by an unbinding process of bound sites, and two or more detached domains may merge into a single larger one.

In principle a detached domain may also split into two domains when an unbound segment of the interface touches the substrate at the interior of the domain (marked by the arrow in Fig.~\ref{fig:reattach}). Let us first consider those regions of the parameter space where such splitting processes are highly suppressed. For example, this is the case when an unbound interface tends to move away from the substrate while it is held bound to it by some short range attractive interaction. In this case the dynamics of the interface may very well be described by a directed percolation process in which the active sites are those bound to the substrate. In fact, in some models~\cite{AlonEtAl96,AlonEtAl98,GinelliEtAl03} it was observed that the depinning transition belongs to the DP universality class. However, the observation of first-order transitions under similar conditions~\cite{HinrichsenEtAl00} suggests the existence of effective long range interactions between the pinning sites, which may provide a mechanism that can lead to first-order transitions in one-dimensional models \cite{Hinrichsen00b}. To explain this mechanism, Ginelli \etal~\cite{GinelliEtAl05} introduced a variant of directed percolation with long-range interactions called $\sigma$-process, which will be described below.

On the other hand, in those regions of the parameter space where the splitting of unbound segments by reattachment in the middle is likely, a different long-range mechanism is needed to describe the dynamics of the contact points. On the level of a DP-like process such long-range interactions can be modeled by \textit{truncated} L{\'e}vy flights, called $\alpha$-process, which will be discussed in Sect.~\ref{sub:alpha}.

Both generalizations, the $\sigma$- and the $\alpha$-process, differ significantly from DP with L{\'e}vy-distributed interactions and waiting times discussed in the previous section. As will be shown, they give rise to fractional powers of the density field (instead of fractional derivatives) in the corresponding Langevin equation.

\subsection{Local transport with non-local rate dependence: The $\sigma$-process}
\label{sub:sigma}

As ordinary DP, the $\sigma$-process involves spontaneous recovery $1\to 0$ and infection of nearest neighbors $10/01 \to 11$ and thus can be considered as a local process. However, in contrast to DP, where the rates for infection and recovery are constant, the infection rate now depends on the actual distance to the nearest neighbor, i.e., it varies with the size of the unbound segment in the corresponding unbinding transition, introducing an effective long-range interaction. More specifically, the infection process may be implemented as
\begin{eqnarray*}
&1  \rightarrow 0  &\mbox{ with rate } 1\,, \\
&0^{\ell}1 \hspace{0.25cm} [10^{\ell}] \rightarrow 0^{\ell -1}11 \hspace{0.25cm} [110^{\ell -1}]
&\mbox{ with rate } \bar{\lambda}(\ell)\,,
\end{eqnarray*}
where $\bar{\lambda}(\ell)=\lambda(1+a/\ell^\sigma)$ depends algebraically on the size $\ell$ of an inactive island of size~$\ell$. The process is controlled by three parameters, namely, the mean infection rate $\lambda$, the control exponent $\sigma$, and an amplitude $a$ controlling the intensity of the long-range interactions. For $a=0$ the model reduces to the ordinary contact process with a transition at $\lambda_c = 1.64892(8)$ which belongs to the DP universality class~\cite{MarroDickman99}. \\

\noindent{\bf \small Mean field approximation:}\\[1mm]
%
Denoting by $\rho(t)$ the average density of active sites at time $t$, the mean field equation in the thermodynamic limit $L \to \infty$ takes the form
\begin{equation}
\frac{d \rho}{d t} =
(\lambda-1)\rho -\lambda \rho^2 + \lambda a \rho^2 \sum_{\ell=1}^{\infty}
\frac{(1-\rho)^\ell}{\ell^{\sigma}} \,.
\label{MFsigma}
\end{equation}
For $\sigma > 1$ the sum on the r.h.s. is finite in the limit $\rho \to 0$ and its contribution amounts to a renormalization of the coefficient of the $\rho^2$ term so that the mean field behavior of a standard DP process with short range interaction is recovered, provided that $a$ is small enough. On the other hand, for $0<\sigma<1$, the leading contribution of the sum can be captured by replacing it with an integral over $d\ell$ which, to leading order in $\rho$, reduces to
\begin{equation}
\partial_t\rho = (\lambda-1)\rho +
a \lambda \Gamma(1-\sigma)\rho^{1+\sigma}-\lambda \rho^2 \,.
\label{MFsigma3}
\end{equation}
Here $\Gamma$ is the standard Gamma function. The leading nonlinear term in this equation involves a non-integer power of the density field, reflecting the long range nature of the interactions. Since its coefficient is positive,  Eq.~(\ref{MFsigma3}) does not admit fixed point solutions for arbitrarily small densities, so that the transition to the absorbing state becomes discontinuous. \\

\begin{figure}
\centering\includegraphics[width=105mm]{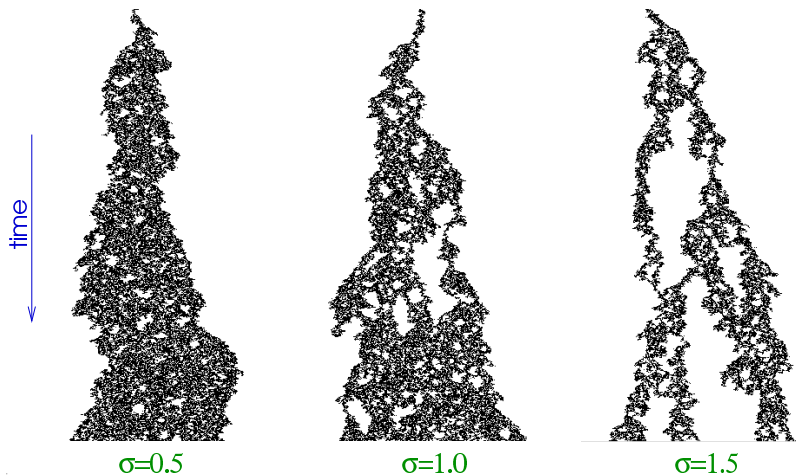}
\caption{\small
Snapshots of the $\sigma$-process close to criticality starting with a single seed for various values of $\sigma$.
\label{fig:sigma}}
\end{figure}

\noindent{\bf \small Numerical simulation}\\[1mm]
%
Numerical simulations suggest that the mean field prediction (DP for $\sigma>1$, first-order transition for $\sigma<1$) is qualitatively correct even in the full model in 1+1 dimensions~\cite{GinelliEtAl05}. To detect the crossover from a continuous to a discontinuous transition in one spatial dimension, it is useful to perform simulations starting with a single seed and to generate a cluster of active sites at criticality. As usual, DP behavior manifests itself by sparsely occupied clusters. Contrarily, first-order transitions can be recognized by the emergence of a single compact domain. As there is no surface tension in one-dimensional systems, these domains fluctuate without bias, hence the process becomes equivalent to a process called compact directed percolation~\cite{Essam89}. In fact, as illustrated in Fig.~\ref{fig:sigma}, for $\sigma<1$ the generated clusters become compact on large scales.

Numerically it is easy to identify these two phases since the exponents of DP in 1+1 dimensions ($\alpha\approx 0.159$, $\theta\approx 0.313$, and $z\approx 1.58$) differ significantly from those of compact directed percolation ($\alpha=1/2$, $\theta=0$, and $z=2$). The simulations carried out in Ref.~\cite{GinelliEtAl05} clearly indicate that the crossover takes place at $\sigma$ close to $1$.

\subsection{Truncated L{\'e}vy flights: The $\alpha$-process}
\label{sub:alpha}

If the parameters of the wetting process are chosen in such a way that an unbound segment of the interface may reattach to the substrate in the interior (marked by the arrow in Fig.~\ref{fig:reattach}), an effective model for the dynamics of binding sites at the bottom layer has to incorporate an additional type of long-range interactions. In numerical studies of wetting transitions, the probability of reattachment at the critical point of the unbinding transition was found to decay algebraically with the distance from the nearest binding site. This motivated Ginelli \etal to study a model called $\alpha$-process which mimics these reattachments\cite{GinelliEtAl06}. The model evolves by random sequential updates with the transition rates
\begin{eqnarray}
\label{process1}
1  &\rightarrow& 0      \hspace{8mm}\mbox{with rate }\hspace{0.5cm} 1 \,, \\
\label{process3}
0 &\rightarrow& 1   \hspace{7mm}\mbox{with rate }\hspace{0.5cm} q/l^\alpha\nonumber \,.
\end{eqnarray}
The rate at which such a process takes place decays algebraically with the distance $l$ from the nearest active site (measured in lattice site units), reflecting the long range nature of the interaction. Note that in contrast to the $\sigma$-model, where infections are local (although taking place with a non-locally determined rate), the $\alpha$ process admits infections over long distances. Obviously, it reduces to ordinary DP in the limit $\alpha \to \infty$.

It should be emphasized that the $\alpha$-process differs significantly from spreading processes with L{\'e}vy-distributed interactions discussed in the previous section. Since an inactive site can only be activated by the \textit{nearest} active site, the interaction range is effectively cut off by half the actual size of the corresponding island of inactive sites. This is to say that infections in the $\alpha$-process are mediated by \textit{truncated} L{\'e}vy flights, which cannot overtake neighboring active sites. \\

\noindent{\bf \small Mean field approximation:}\\[1mm]
%
Assuming different sites to be uncorrelated, the probability of an inactive site to lie at a distance $l$ from the nearest active site is $(1-(1-\rho)^2) (1-\rho)^{2 l -2}$. Summing up the contributions for all distances $l$, the mean-field equation governing the dynamics of the density of active sites takes the form
\begin{equation}
\partial_t\rho(t)=-\rho(t) + q \rho(t) (2-\rho(t)) \sum_{l=1}^{\infty}
\frac{(1-\rho(t))^{2l-1}}{l^{\alpha}} \,.
\label{MFalpha1}
\end{equation}
Turning the sum into an integral the leading terms in $\rho(t)$ in this equation can be calculated. It is found that for $\alpha<1$ and, to leading order in $\rho(t)$, the mean field equation turns into $\partial_t\rho(t)=r\rho(t) + u \rho(t)^{\alpha}$,  where $r$ and $u$ are constants with $u>0$. Since the leading term in this equation is $\rho^{\alpha}$, and since its coefficient is positive, the absorbing state $(\rho=0)$ is always unstable and no transition takes place for any finite $q$. For $1<\alpha<2$, however, the leading term in Eq.~(\ref{MFalpha1}) is the linear one and the coefficient $u$ is negative. This results in a continuous transition different from DP which takes place at $r=0$ corresponding to a non-vanishing rate $q$.

The above mean field equation has to be extended by a term that describes the transport mechanism. In systems with non-conserved short-range interactions this term is a Laplacian, while in systems with unrestricted L{\'e}vy-distributed interactions it would be a fractional derivative (see Eq.~(\ref{eq:SpatialLevyOperator})). In the $\alpha$-process, however, the interactions are power-law distributed but their range is cut off at the distance to the closest active site. Therefore, in the large-scale limit (which the mean field equations are supposed to describe) the interactions are effectively short-ranged and thus described by a Laplacian. Instead the effect of the non-local infection process is to make the diffusion coefficient $D$ dependent on the density $\rho$. As shown in Ref.~\cite{GinelliEtAl06}, for $1 < \alpha < 2$ this reasoning leads to the mean field equation
\begin{equation}
\label{MFimproved}
\partial_t\rho(\rvec,t)=r\rho(\rvec,t) + u[\rho(\rvec,t)]^\alpha + [\rho(\rvec,t)]^{\alpha-3}\nabla^2\rho(\rvec,t)
\end{equation}
for which dimensional analysis yields the complete set of mean-field exponents
\begin{equation}
\nu_\perp^{\rm \scriptscriptstyle MF} = \beta^{\rm \scriptscriptstyle MF}= \frac{1}{\alpha-1}\,, \quad \nu_\parallel^{\rm \scriptscriptstyle MF}=1\,,\qquad z^{\rm \scriptscriptstyle MF}=\alpha-1\,.
\end{equation}
On the other hand, for $2<\alpha<3$ the leading non-linear term is the quadratic one, leading to the lowest-order mean field equation
\begin{equation}
\label{MFimproved2}
\partial_t\rho(\rvec,t)=r\rho(\rvec,t) + v[\rho(\rvec,t)]^2 + [\rho(\rvec,t)]^{\alpha-3}\nabla^2\rho(\rvec,t).
\end{equation}
with the critical exponents
\begin{equation}
\nu_\perp^{\rm \scriptscriptstyle MF} = 2 - \frac{\alpha}{2}\,, \quad \beta^{\rm \scriptscriptstyle MF}= \nu_\parallel^{\rm \scriptscriptstyle MF}=1\,,\qquad z^{\rm \scriptscriptstyle MF}=2/(4-\alpha)\,.
\end{equation}
Finally, for $\alpha>3$ the leading-order mean field equation is the same as the one of DP, hence we expect the process to cross over to ordinary directed percolation with $z=2$. This phase structure is also observed numerically, although with different values of the critical exponents~\cite{GinelliEtAl06}.\\

\noindent{\bf \small Combination of the $\alpha$- and the $\sigma$ process:}\\[1mm]
%
Both models, the $\sigma$-process and the $\alpha$-process, describe different aspects of the effective dynamics of binding sites in a non-equilibrium wetting process. In most situations, however, one expects a mixture of both mechanisms. Such a combined $\alpha-\sigma$ process may be modeled by the transitions
\begin{eqnarray}
\label{combprocess1}
1  &\rightarrow & 0      \hspace{8mm}\mbox{with rate }\hspace{0.5cm} 1\nonumber \\
\label{combprocess2}
10,01 &\rightarrow & 11  \hspace{5mm}\mbox{with rate }\hspace{0.5cm} q(1+a/m^\sigma) \\
\label{combprocess3}
000 &\rightarrow & 010   \hspace{7mm}\mbox{with rate }\hspace{0.5cm} bq/l^\alpha \nonumber
\end{eqnarray}
where $m$ is the size of the inactive island and $l$ denotes the distance from the nearest active site. Here $q$ is the control parameter, $\alpha$ and $\sigma$ are the control exponents, and $a$ and $b$ are constants. The ``pure'' $\alpha$- and $\sigma$-processes are recovered in the limits $\sigma \to \infty$ and $\alpha \to \infty$, respectively.

\begin{figure}
\centering\includegraphics[width=55mm]{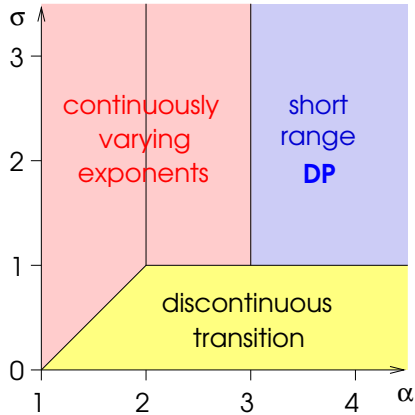}
\caption{\small
Phase diagram of the combined $\alpha$-$\sigma$ process
\label{fig:alphasigma} }
\end{figure}

Within a mean field approach, this combined $\alpha$-$\sigma$ process is described by the equation
\begin{equation}
\partial_t\rho \;=\;
-\rho \,+\,
q \rho^2 \sum_{m=1}^{\infty} (1+\frac{a}{m^{\sigma}})(1-\rho)^m  \,+\,
b q \rho (2 - \rho) \sum_{l=2}^{\infty} \frac{(1-\rho)^{2l-1}}{l^{\alpha}} \,.
\label{MFsigmaalpha}
\end{equation}
In the regime of interest, namely for $0<\sigma<1$ and $1<\alpha<2$, the mean-field equation can be written to leading order as
\begin{equation}
\label{MFSigmaAlpha}
\partial_t\rho=r\rho + p\rho^{1+\sigma} - u\rho^\alpha + 0(\rho^2)\,,
\end{equation}
where
\begin{equation}
r = q-1+bq{2^{2-\alpha}\over{\alpha-1}}\,, \quad
p = aq\,\Gamma(1-\sigma)\,,\quad
u = bq\,{{2^{\alpha-1}}\over{\alpha-1}}\Gamma(2-\alpha)
\end{equation}
are $q$-dependent constants. As shown in Ref.~\cite{GinelliEtAl06}, the corresponding mean field diagram (see Fig.~\ref{fig:alphasigma}) contains four phases, namely,
\begin{enumerate}
\item [(i)]a DP phase for $\sigma>1,\alpha>3$,

\item [(ii)] a phase for $\sigma> \alpha -1$ and $1<\alpha < 2$, where the transition is second order with continuously varying exponents.

\item [(iii)] a phase for $\sigma > 1$ and $2< \alpha < 3$, where the transition is second order but where $z$ is the only continuously varying exponent.

\item [(iv)] a phase for $\sigma < \alpha - 1$ and $0< \sigma < 1$, where the transition is discontinuous.
\end{enumerate}

\section{Concluding remarks}

Continuous phase transitions into absorbing states continue to fascinate theoretical physicists because of their universal properties which are determined by symmetries and conservation laws rather than microscopic details. On a field-theoretic level the origin of universality can be understood in terms of relevant and irrelevant operators in the corresponding action. Usually it is assumed that the process evolves by a Markovian dynamics with local interactions, as described to leading order by differential operators such as $\partial_t$ and $\nabla^2$. Generalizing these models by introducing long-range interactions and non-Markovian effects with power-law characteristics these universal properties may change.

In Sect.~\ref{sec:InteractingLevy} we have discussed how \textit{unrestricted} long-range interactions over a distance~$r$ distributed as $P(r)\sim r^{-d-\sigma}$ and waiting times $\Delta t$ with a power-law distribution $P(\Delta t)\sim t^{-1-\kappa}$ affect phase transition belonging to the universality class of directed percolation. Generically one observes the following phenomenology: If the control exponent $\sigma$ (or $\kappa$) is larger than a certain upper threshold $\sigma^*$ (or $\kappa^*$) the model behaves effectively as if the interactions where short-ranged. If the exponent is smaller than a certain lower threshold $\sigma_c$ (or $\kappa_c$), the couplings become so long-ranged that an effective mean-field behavior is observed. In between there is a non-trivial phase in which the critical exponents $\beta$, $\nu_\perp$, and $\nu_\parallel$ vary continuously with the control exponent. Generally long-range interactions in space as well as in time tend to reduce fluctuation effects, thereby decreasing the upper critical dimension of the process.

On a field-theoretical level L{\'e}vy-distributed long range interactions can be studied by adding terms with fractional derivatives. Generalizing directed percolation in this way one obtains a renormalizable field theory describing a spectrum of universality classes parameterized by $\sigma$ and $\kappa$. Moreover, it is found that the additional terms involving fractional derivatives are not renormalized, implying exact scaling relations among the critical exponents in the non-trivial phase.

These scaling relations can be used to compute the upper thresholds $\sigma^*$ and $\kappa^*$, where the crossover to the short-range behavior takes place. Surprisingly one finds that $\sigma^*>2$ and $\kappa^*>1$. This example demonstrates that in an interacting field theory the thresholds of $\sigma$  and $\kappa$, where the fractional derivatives $\tilde{\nabla}^\sigma$ and $\tilde{\partial}_t^\kappa$ cross over to their short-range counterparts $\nabla^2$ and $\partial_t$, may be shifted away from the naive thresholds $\sigma=2$ and $\kappa=1$ that would be obtained if these operators were acting on smooth functions. More recently the same issue was debated in the context of the Ising model with L{\'e}vy-distributed couplings~\cite{FisherEtAl72,Luijten98a,LuijtenBloete02} without being aware that the analogous problem was already solved for DP.

In Sect.~\ref{sec:Restricted} we have discussed a different type of long range interactions motivated by recent studies of non-equilibrium wetting processes. In the first case, called $\sigma$-process, the interactions are local but the coupling constant has a non-local dependence. This modification induces a crossover from directed percolation to a first-order transition. In the second case, called $\alpha$-process, the interactions are non-local and power-law distributed but in contrast to the unrestricted interactions discussed in Sect.~\ref{sec:InteractingLevy} they are now cut off at the actual distance to the nearest particle. It turns out that fractional derivatives are no longer suitable to describe such truncated L{\'e}vy flights, instead one gets fractional powers of the density field and an algebraically varying diffusion constant in the corresponding Langevin equation.

Obviously, the study of reaction-diffusion models and non-equilibrium critical phenomena with long-range interactions is still at its beginning. As demonstrated in this review, such problems are very fascinating as they can be treated both by numerical and analytical methods. Since long-range transport mechanisms are frequently observed in complex systems, it is the hope that the theoretical insight from simple models will be useful for the understanding of experiments in the future.

\noindent\textbf{Acknowledgement}

\noindent
I would like to thank the Isaac Newton Institute in Cambridge for hospitality, where parts of this work were done.

\newpage
\appendix
\section{}
{\bf \noindent Field-theoretic approach to directed percolation with L{\'e}vy flights}\\

\noindent
This appendix sketches the field-theoretic approach to directed percolation with spatial L{\'e}vy flights. This problem was first solved by Janssen et. al.~\cite{JanssenEtAl99} using dimensional regularization techniques. Here we present an alternative calculation using Wilsons momentum shell approach, leading to the same results.

Starting point is the field-theoretic action
\begin{eqnarray}
\label{eq:AppendixAction}
S[\bar\psi,\psi] &=& \int {\rm d}^dr \, {\rm d}t \Bigl\{
\bar\psi(\tau \partial_t - a - D \nabla^2 - \tilde{D} \tilde{\nabla}^{\sigma} )\psi \\
&&\hspace{20mm}
+g \psi \bar{\psi} (\psi-\bar\psi)- \rho_0 \delta(t) \bar\psi\nonumber
\Bigr\}\,.
\end{eqnarray}
Here $\tau$ is a coefficient fixing the time scale, $a$ plays the role of the reduced percolation probability, $D$ and $\tilde{D}$ are the coefficients for short-range diffusion and L{\'e}vy flights, $g$ is the symmetrized nonlinear coupling constant, and $\rho_0$ denotes the initial density of particles at $t=0$. The only properties of the fractional derivative $\tilde{\nabla}^{\sigma}$ used in the following renormalization group calculation are its scaling dimension as well as its action in momentum space
\begin{equation}
\tilde{\nabla}^{\sigma} e^{i \kvec\cdot\rvec} \;=\;  -|\kvec|^\sigma e^{i \kvec\cdot\rvec}\,.
\end{equation}
%
{\bf Scaling transformation:}\\[3mm]
%
The first step of Wilsons renormalization group procedure is to rescale the action by
\begin{equation}
\label{eq:AppendixRescaling}
\rvec \to b \rvec  ,\quad
t \to b^z t  ,\quad
\rho_0  \to b^{-d_{\rho_0}}\rho_0,\quad
\psi  \to b^{-d_\psi} \psi  ,\quad
\bar\psi  \to b^{-d_{\bar\psi}} \bar\psi,
\end{equation}
where $b$ is a dilatation parameter, $z=\nu_\parallel/\nu_\perp$ is the dynamical critical exponent, while $d_\psi$, $d_{\bar\psi}$, and $d_{\rho_0}$ are scaling dimensions associated with the fields. Because of the well-known rapidity reversal symmetry of DP, which holds also in the presence of L{\'e}vy flights, the fields $\psi$ and $\bar\psi$ scale identically, meaning that  $d_{\bar\psi} = d_\psi$. It is convenient to express these scaling dimensions in terms of the anomalous dimensions $\eta_\psi$ and $\eta_{\rho_0}$ by
\begin{equation}
d_\psi = \frac12 ( d+\eta_\psi) \,, \qquad d_{\rho_0}= \frac12(d+\eta_{\rho_0})\,.
\end{equation}
The scaling transformation~(\ref{eq:AppendixRescaling}) leads to a change of the coefficients in the action~(\ref{eq:AppendixAction}). For an infinitesimal dilatation $b=1+\ell$ these changes are given to lowest order by
\begin{eqnarray}
\partial_\ell \ln \tau 		&=& -\eta_\psi \nonumber \\
\partial_\ell \ln a 		&=& z-\eta_\psi \nonumber \\
\partial_\ell \ln D 		&=& z-\eta_\psi-2 \nonumber \\
\partial_\ell \ln \tilde{D} 	&=& z-\eta_\psi-\sigma  \\
\partial_\ell \ln g 		&=& z-3\eta_\psi/2 - d/2 \nonumber \\
\partial_\ell \ln \rho_0 	&=& -(\eta_\psi+\eta_{\rho_0})/2 \nonumber
\end{eqnarray}
These are the renormalization group equations on tree level which are expected to be valid above the upper critical dimension $d_c=2\sigma$, where loop corrections are irrelevant. The fixed point of these equations is $z^{*}=2$ and $\eta^{*}_{\psi}=\eta^{*}_{\rho_0}=0$. Together with $\beta^{MF}=1$ and $d_\psi=\beta/\nu_\perp$ they determine the mean field exponents
\begin{equation}
\beta^{MF}=1\,,\quad \nu_\perp^{MF} = \sigma^{-1}\,,\quad \nu_\parallel^{MF}=1\,
\end{equation}
which are expected to be exact above the upper critical dimension $d_c=2\sigma$. Moreover, it turns out that the critical initial slip exponent
\begin{equation}
\theta = \frac{\eta_{\rho_0}-\eta_\psi}{2z}\,,
\end{equation}
which describes the average increase of the particle number in simulations starting with a single seed, vanishes at tree level.\\

\begin{figure}
\centering\includegraphics[width=85mm]{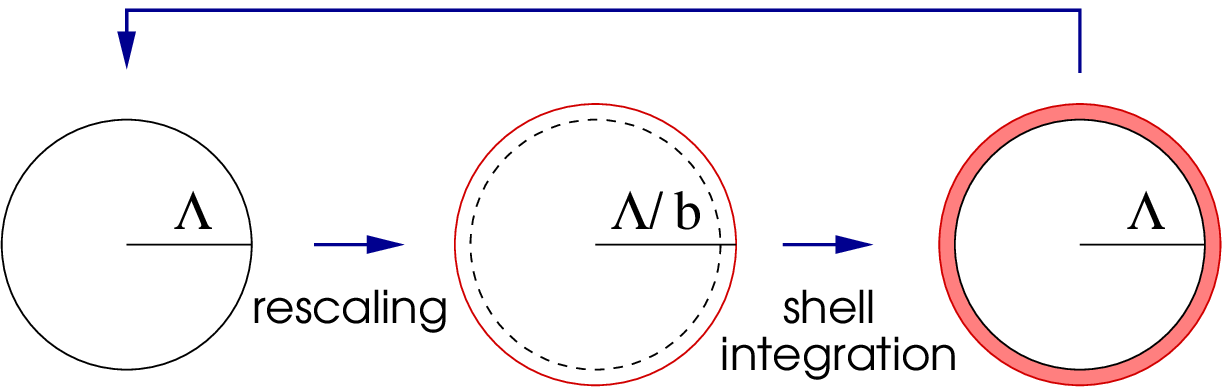}
\caption{\small
Schematic sketch of Wilsons renormalization group procedure. A minimal length representing the lattice spacing is introduced by imposing an upper cutoff $\Lambda$ in momentum space. In a first step all quantities are rescaled infinitesimally, modifying the cutoff scale. In order to compensate this modification, the high-wavenumber modes within the momentum shell are integrated out in a second step and the resulting contributions are absorbed to lowest order in the coefficients of the action.
\label{fig:wilson} }
\end{figure}

\noindent{\bf \small One-loop diagrams:}\\[3mm]
%
Wilsons renormalization group approach computes loop corrections by introducing an upper cutoff $|\kvec| \leq \Lambda$ in momentum space. Since the rescaling procedure changes the cutoff by $\Lambda\to b^{-1} \Lambda$ a second step is required that sends the cutoff back to its original value (see Fig.~\ref{fig:wilson}). This can be done integrating out all fast oscillating modes within the (infinitesimal) momentum shell $\Lambda < |\kvec| < b^{-1}\Lambda$ and then to absorb the resulting contributions to lowest order into the coefficients of the action.

In Wilsons renormalization group approach to ordinary DP (see e.g. Ref.~\cite{WijlandEtAl98}) the propagator in momentum space is renormalized to one-loop order by
\begin{equation}
- G_0^{-1}(\kvec,\omega)^{\rm ren.} = - G_0^{-1}(\kvec,\omega) - \frac{g^2}{2} I_P
\end{equation}
where
\begin{equation}
\label{eq:FreePropagator}
G_0=(Dk^2-a-i\tau\omega)^{-1}
\end{equation}
is the free propagator and $I_p$ denotes the loop integral
\begin{equation}
\label{Appendix:IP}
I_P(\kvec,\omega)=\int_> \frac{{\rm d}^dk'{\rm d}\omega'}{(2\pi)^{d+1}}\;
G_0\biggl(\frac{\kvec}{2}+\kvec',\frac{\omega}{2}+\omega'\biggr)\,
G_0\biggl(\frac{\kvec}{2}-\kvec',\frac{\omega}{2}-\omega'\biggr)\,.
\end{equation}
Here the symbol `$>$' indicates that the integration is restricted to the momentum shell $\Lambda < |\kvec'| < \Lambda/b$. Similarly, the non-linear vertices are renormalized by
\begin{equation}
-g^{\rm ren.} = -g + 2 g^3 I_V
\end{equation}
with the loop integral
\begin{equation}
\label{Appendix:IV}
I_V = \int_> \frac{{\rm d}^dk{\rm d}\omega}{(2\pi)^{d+1}}\;
G_0^2(\kvec,\omega)\, G_0(-\kvec,-\omega)\,.
\end{equation}
As discussed in the main text, spatial L{\'e}vy flights can be introduced by replacing the free propagator~(\ref{eq:FreePropagator}) with
\begin{equation}
\label{eq:ModifiedFreePropagator}
G_0=(Dk^2+\tilde{D}|k|^\sigma-a-i\tau\omega)^{-1}
\end{equation}
while the structure of the loop integrals and their multiplicities do not change. Therefore, all what has to be done is to re-evaluate the loop integrals~(\ref{Appendix:IP}) and ~(\ref{Appendix:IV}). As explained above, it is important to retain the short-range term $k^2$ in the action since the L{\'e}vy operator does not renormalize itself, instead it renormalizes the Laplacian.\\

\noindent{\bf \small Calculation of the loop integrals}\\[3mm]
%
Let us first evaluate the propagator loop integral~(\ref{Appendix:IP}). With the abbreviation
\begin{equation}
A^\pm = D (\kvec'\pm\kvec/2)^2 + \tilde{D}|\kvec'\pm\kvec/2|^\sigma - a- i\tau\omega/2
\end{equation}
this integral can be expressed as
\begin{eqnarray}
I_P(\kvec,\omega) &=& \int_> \frac{{\rm d}^dk'{\rm d}\omega'}{(2\pi)^{d+1}}\;
\frac{1}{(A^+-i\tau\omega')(A^--i\tau\omega')} \nonumber \\
&=& \frac{1}{(2\pi)^d \tau} \, \int_> {\rm d}^dk' \, \frac{1}{A^++A^-}\,.
\end{eqnarray}
Since $\kvec'$ is much larger than $\kvec$ we may expand the denominator by
\begin{eqnarray}
&&|\kvec'+\frac{\kvec}{2}|^\sigma+|\kvec'-\frac{\kvec}{2}|^\sigma =\\
&&=2|\kvec'|^\sigma + \frac{\sigma}{4}|\kvec'|^{\sigma-2}k^2
+\frac{\sigma}{2}(\frac{\sigma}{2}-1)\,|\kvec'|^{\sigma-2}k^2\cos^2\theta + \mathcal{O}(k^4)\nonumber
\end{eqnarray}
where $\theta$ is the angle between $\kvec$ and $\kvec'$. The loop integral over the momentum shell at $|\kvec'|\approx \Lambda$ with the thickness $\ell$ is given by
\begin{eqnarray}
I_P(\kvec,\omega)   &=&  \frac{\ell \Lambda^d}{(2\pi)^d\tau} S_{d-1}
\int_0^\pi {\rm d}\theta \, \sin^{d-2} \theta
\Bigl(
2 D \Lambda^2 + \frac{D}{2} k^2 + 2\tilde{D} \Lambda^\sigma + \frac{\tilde{D} \sigma}{4}\Lambda^{\sigma-2}k^2 \nonumber\\
&& \hspace{20mm} + \frac{\tilde{D}\sigma}{4}(\sigma-2)\Lambda^{\sigma-2}k^2\cos^2\theta-2a-i\tau\omega
\Bigr)^{-1}\,,
\end{eqnarray}
where $S_d=2 \pi^{d/2}/\Gamma(d/2)$ is the surface of $d$-dimensional unit sphere. Carrying out the integration one obtains
\begin{equation}
I_P \;\simeq\; \frac{\ell \Lambda^d K_d}{2 \tau}
\left(
\frac{1}{\mathcal{N}} -
\frac{D+\frac{\tilde{D}\sigma}{2}\Lambda^{\sigma-2}
+ \frac{\tilde{D}\sigma}{d}(\frac{\sigma}{2}-1)\Lambda^{\sigma-2}}{4\mathcal{N}^2} \, k^2 +
\frac{i \tau}{2 \mathcal{N}^2}\omega
\right),
\end{equation}
where $K_d=S_d/(2\pi)^d$ and $\mathcal{N}=D \Lambda^2 +\tilde{D}\Lambda^\sigma-a$.
Thus, integrating out the one-loop corrections within the momentum shell one obtains to lowest order three terms which have the same structure as the terms in the free part of the action. Similarly we can compute the vertex loop integral, giving
\begin{eqnarray}
I_V  &=&  \int_> \frac{{\rm d}^dk{\rm d}\omega}{(2\pi)^{d+1}}\;
G_0^2(\kvec,\omega) G_0(-\kvec,-\omega) \\
&=&  \int_> \frac{{\rm d}^dk{\rm d}\omega}{(2\pi)^{d+1}}\;
\frac{1}{(Dk^2+\tilde{D}|\kvec|^\sigma-a-i\tau\omega)^2\,(Dk^2+\tilde{D}|\kvec|^\sigma-a-i\tau\omega)} \nonumber\\
 &=& \frac{1}{\tau} \int_> \frac{{\rm d}^dk}{(2\pi)^{d}}\;
\frac{1}{4(D k^2+\tilde{D}|\kvec|^\sigma-a)^2} \nonumber
 \;=\; \frac{\ell K_d \Lambda^d}{4 \tau \mathcal{N}^2}\,.
\end{eqnarray}
\vspace{4mm}

\noindent{\bf \small Renormalization group flow equations}\\[3mm]
%
Absorbing the contributions from the shell integrals in the coefficients of the action, the renormalization group equations to one-loop order are given by
\begin{eqnarray}
\partial_\ell \ln \tau 		&=& -\eta_\psi -2S_2\nonumber \\
\partial_\ell \ln a 		&=& z-\eta_\psi-S_1 \nonumber \\
\partial_\ell \ln D 		&=& z-\eta_\psi-2
-S_2\left( 1-\frac{1}{D}\biggl(\frac{\sigma}{2}+\frac{\sigma^2}{2d}-\frac{\sigma}{d}\biggr)\tilde{D}\Lambda^{\sigma-2}\right) \nonumber \\
\partial_\ell \ln \tilde{D} 	&=& z-\eta_\psi-\sigma \quad \mbox{\begin{tiny}(no loop corrections)\end{tiny}} \\
\partial_\ell \ln g 		&=& z-3\eta_\psi/2 - d/2 -8S_2\nonumber \\
\partial_\ell \ln \rho_0 	&=& -(\eta_\psi+\eta_{\rho_0})/2\quad\mbox{\begin{tiny}(no loop corrections)\end{tiny}} \nonumber
\end{eqnarray}
where
\begin{equation}
S_1=\frac{g^2\Lambda^dK_d}{4a\tau\mathcal{N}}\,,\quad
S_2=\frac{g^2\Lambda^dK_d}{16\tau\mathcal{N}^2}\,.
\end{equation}
As discussed in the main text, the coefficients $\tilde{D}$ and $\rho_0$ are not renormalized by loop corrections to any order of perturbation theory, implying the well-known generalized hyperscaling relation for the critical initial slip exponent
\begin{equation}
\theta = \frac{\eta_{\rho_0}-\eta_\psi}{2z} = \frac{d}{z} - 2\delta
\end{equation}
as well as a new scaling relation among the three standard exponents
\begin{equation}
\nu_\parallel-\nu_\perp(\sigma-d)-2\beta = 0\,.
\end{equation}
%
\noindent{\bf \small Fixed point and critical exponents}\\[3mm]
%
Let us now assume that the coefficients $\tilde{D}$ and $\tau$ are constant during renormalization. Denoting by $\epsilon=d_c-d=2\sigma-d$ the difference from the upper critical dimension, the remaining RG flow equations take the form
\begin{eqnarray}
\partial_\ell \ln a 		&=& \sigma-S_1 \nonumber \\
\partial_\ell \ln D 		&=& \sigma-2
-S_2\left( 1-\frac{1}{D}\Bigl(\frac{\sigma}{2}+\frac{\sigma^2}{2d}-\frac{\sigma}{d}\Bigr)\tilde{D}\Lambda^{\sigma-2}\right) \nonumber \\
\partial_\ell \ln g 		&=& \epsilon/2 - 7S_2\nonumber
\end{eqnarray}
These equations have the fixed point
\begin{equation}
S_1^* = \sigma\,,\qquad S_2^* = \epsilon/4\,,
\end{equation}
or in terms of the original coefficients
\begin{eqnarray}
a^* &=& \frac{2 \tilde{D} \Lambda^\sigma}{7\sigma} \epsilon + \mathcal{O}(\epsilon^2) \nonumber \\
({g^*})^2  &=& \frac{8 \tau \tilde{D}^2 \Lambda^{2\sigma-d}}{7K_d}\epsilon  + \mathcal{O}(\epsilon^2) \\
D^*  &=& \frac{\tilde{D} \Lambda^{\sigma-2}\biggl(\frac{\sigma}{2}+\frac{\sigma^2}{2d}-\frac{\sigma}{d}\biggr)}{14(2-\sigma)}\epsilon + \mathcal{O}(\epsilon^2) \nonumber \,.
\end{eqnarray}
In the vicinity of the fixed point the linearized RG flow is governed by the matrix
\begin{equation}
M=\left(
\begin{array}{cc}
\partial_a a(\sigma-S_1) & \partial_a a(\epsilon/2-7D_2) \\
\partial_g g (\sigma-S_1) & \partial_g g (\epsilon/2-7D_2)
\end{array}
\right)_{g=g^*; \;a=a^*}
\end{equation}
To first order in $\epsilon$, the eigenvalues of this matrix are $\alpha_1=\sigma-2\epsilon/7$ and $\alpha_2=-\epsilon$. Here the first eigenvalue is positive and represents the attractive line of the renormalization group flow, meaning that $a$ vanishes as $\Lambda^{-\alpha_1}$ as the fixed point is approached. Since $a$ plays the role of the reduced percolation probability, we may identify $\nu_\perp$ with $\alpha_1^{-1}$. Having reached the fixed point, the postulated stationarity of $\tilde{D}$ and $\tau$ implies that $\eta_\psi=-\epsilon/7$ and $z=\sigma-\epsilon/7$. Expressing these identities in terms of the standard exponents one is led to the results in Eq.~(\ref{eq:SpatialCriticalExponents}).\\

\newpage
\noindent\textbf{\large References}
\vspace{3mm}

\bibliographystyle{iopart-num}
\bibliography{/home/hinrichsen/Dateien/Literatur/master}
\end{document}